\documentclass[useAMS,usenatbib]{mn2e}

\usepackage{graphicx}
\usepackage{scalefnt}

\newif\ifAMStwofonts
\bibpunct{(}{)}{;}{a}{}{,}
\newcommand{\vR}{v_{\rm{R}}}
\newcommand{\vphi}{v_{\phi}}
\newcommand{\vz}{v_{\rm{z}}}
\newcommand{\Lz}{L_z}
\newcommand{\vlos}{v_{\rm{los}}}
\title[Simulations of minor mergers. II.]{ Simulations of minor
mergers. II. The phase-space structure of thick discs}
\author[\'A. Villalobos and A. Helmi]{\'Alvaro
Villalobos\thanks{E-mail: villalobos@astro.rug.nl} and Amina
Helmi\thanks{E-mail: ahelmi@astro.rug.nl}\\ Kapteyn Astronomical
Institute, University of Groningen, P.O. Box 800, 9700 AV Groningen,
The Netherlands\\}
\begin{document}

%%\date{Accepted 1988 December 15. Received 1988 December 14; in original 
%%form 1988 October 11}
%%\pagerange{\pageref{firstpage}--\pageref{lastpage}} \pubyear{2002}

\date{}
\maketitle

%%\label{firstpage}

\begin{abstract}
We analyse the phase-space structure of simulated thick discs that are
the result of a significant merger between a disc galaxy and a
satellite.  Our main goal is to establish what would be the
characteristic imprints of a merger origin for the Galactic thick
disc.  We find that the spatial distribution predicted for thick disc
stars is asymmetric, seemingly in agreement with recent observations
of the Milky Way thick disc. Near the Sun, the accreted stars are
expected to rotate more slowly, to have broad velocity distributions,
and to occupy preferentially the wings of the line-of-sight velocity
distributions. The majority of the stars in our model thick discs have
low eccentricity orbits (in clear reference to the pre-existing heated
disc) which gives rise to a characteristic (sinusoidal) pattern for
their line of sight velocities as function of galactic longitude.  The
$z$-component of the angular momentum of thick disc stars provides a
clear discriminant between stars from the pre-existing disc and those
from the satellite, particularly at large radii. These results are
robust against the particular choices of initial conditions made in
our simulations, and thus provide clean tests of the disc heating via
a minor merger scenario for the formation of thick discs.
\end{abstract}


\section{Introduction}
\defcitealias{quinn1986}{QG86}
\defcitealias{quinn1993}{QHF93}
\defcitealias{walker1996}{WMH96}
\defcitealias{huang1997}{HC97}
\defcitealias{velazquez1999}{VW99}
\defcitealias{villalobos-helmi2008}{Paper I}

Thick discs appear to be a common feature of disc galaxies
\citep[][and references therein]{yoachim2006}. They are believed to be
snap-frozen relics of disc galaxy formation that took place at
high-redshift \citep{freeman2002}. This is because their dominant
stellar populations appear to be old, as indicated for example by the
colours of the envelopes of edge-on disc galaxies \citep[see,
e.g.][and follow-up papers]{dalcanton2002}. More direct evidence comes
from local studies of the Milky Way's thick disc, whose stars are
mostly older than 10--12 Gyr \citep[e.g.][]{edvardsson1993}.

In the case of the Milky Way, the thick disc has structural, kinematic
and chemical properties that in general are significantly different
compared to the other Galactic components
\citep{bensby2005,juric2008,veltz2008}. This implies that it is quite
likely it has followed its independent evolutionary path, as discussed
in \citet[]{wyse2004} and references therein \citep[but see
also][]{ivezic2008}.

As stated above, the Galactic thick disc provides us with a window
into the high-redshift Universe. However unlike the Galactic stellar
halo, which is also ancient, the thick disc contains a non-negligible
fraction of the total stellar mass of the Galaxy \citep[between 6 and
15 per cent, e.g.][]{robin2003,juric2008}, what enhances its
importance as a tracer of the events that took place at early
epochs. On the other hand, and like for the stellar halo, imprints of
that early history may be present in the form of dynamical or chemical
substructure. This information could, in principle, be retrieved
relatively easily from large surveys of nearby stars implying that we
can hope to directly test the various scenarios proposed for its
formation.

These various scenarios may be classified according to the relative
importance of dissipative processes. The collapse of a gas cloud with
a large scale-height \citep{eggen1962,burkert1992}, or intense star
formation \citep{kroupa2002}, perhaps triggered by gas rich mergers
\citep{brook2004,bournaud2007a} are two dissipationally driven models.
On the other hand, the vertical heating of a thin disc during a merger
event \citep[e.g.][and references therein]{kazantzidis2008}; or the
direct accretion of satellites proposed by \citet{abadi2003} are
examples of (mostly) dissipationless processes.

In the case of a merger origin of the Galactic thick disc, one may
expect significant substructure to be present, particularly in the
phase-space distribution of its stars. In fact, evidence of such
merger debris in our Galaxy has been mounting over the past
decade. Examples are the substantial group of stars located a few kpc
from the Galactic plane with kinematics intermediate between the
canonical thick disc and the canonical stellar halo \citep{gil2002}; 
or the significant asymmetry in the distribution of thick
disc stars in the first Galactic quadrant with respect to the fourth
\citep{parker03,parker04,larsen08}. In addition to
substructure of the thin/thick disc such as the Arcturus stream 
\citep{eggen1996,navarro2004}, the Monoceros ring \citep{yanny2003},
the Canis Major dwarf \citep{martin2004}, distinctive stellar groups
in the solar vicinity with peculiar ages, metallicities and kinematics
have also been discovered \citep{helmi2006}.  All these observations
thus appear to support models in which the thick disc was formed by
accretion and/or merger events.

In \citet[][hereafter, Paper I]{villalobos-helmi2008} we presented a series of simulations
of minor mergers between a disc galaxy and a relatively massive
satellite, with the aim of modelling the formation of a thick disc
\citep[a scenario first explored thoroughly by][and which has received
much attention since then]{quinn1993}. The present work is intended as
a follow-up study where we particularly focus on characterising the
phase-space properties of these simulated thick discs. Our ultimate
goal is to find sets of observables that would allow us to recover
traces of the merger that may have lead to the formation of the
Galactic thick disc. In principle, it should also be possible to
distinguish stars from the intruder satellite from those of the heated
disc in the final aftermath.

We are also motivated to pursue such a study at this point in time by
the surveys that are currently mapping the Milky Way galaxy and its
components in great detail. In particular, the spectroscopic RAVE and
SEGUE/SDSS surveys are both providing large samples of stars with
accurate kinematics \citep{zwitter2008,sdss-dr6,segue}. Such surveys
will enable a more precise characterisation of the Galactic thick
disc, and should allow us to test the various formation scenarios
discussed above. On a slightly longer timescale, the space astrometric
mission {\em Gaia} will provide full phase-space coordinates for
hundreds of millions of stars \citep{perryman2001}, from which we
should be able to establish how the Galactic thick disc was assembled.

The outline of this paper is the following. In Section 2 we briefly
describe the numerical simulations that are used in this study.  In Section
3 we characterise the velocity distribution in local volumes
resembling the Solar neighbourhood. In this section we put special
emphasis on understanding the predicted distributions of heliocentric
line-of-sight velocities since, in practice, these can be obtained
with high accuracy for large samples of stars, and are available even
at the present time. Finally Section 4 summarises our conclusions.  

\section{Simulations}

\subsection{Description}

In \citetalias{villalobos-helmi2008} we have
performed a series of dissipationless $N$-body simulations of a single
merger between a pre-existing disc galaxy and a satellite, in order to
study the formation of thick discs in a context of disc heating. Among
the main results presented in \citetalias{villalobos-helmi2008}, we
find that such mergers are able to produce thick discs that are both
structurally and kinematically similar to those observed in the Milky
Way and in external galaxies \citep[see also][and references
therein]{velazquez1999}.  Structurally, the simulated thick discs have
larger scale-lengths compared to the initial disc and their
scaleheights are 3--6 times larger with a clear dependence on the
initial inclination of the decaying satellite.  When compared to
observations of the Galactic thick disc, the simulations seem to
favour mergers with low/intermediate initial inclinations for the
formation of this component, as suggested by the measured
$\sigma_{\rm z}/\sigma_{\rm R}$ ratio \citetext{as determined by, e.g.,
\citealt{soubiran2003,vallenari2006}} and from the presence and
amplitude of vertical gradients in the mean rotation
\citep{girard2006,ivezic2008}.

These simulations are therefore a good starting point to study the
detailed phase-space structure of the remnant system, and to establish
if it is possible to dynamically distinguish stars from the
pre-existing disc from those of the accreted satellite. Furthermore,
they may also be used to develop indicators to test the validity of
this particular formation scenario for the Galactic thick disc.

Below we summarise our suite of numerical simulations, but refer the
interested reader to \citetalias{villalobos-helmi2008} for more
details. 

We have explored the following initial configurations for the merger:
(i) the structure and kinematics of the primary disc resemble those of
a) the present Milky Way (``$z$=0'' experiments); or b) a disc at
redshift one (``$z$=1'') according to the model of \citet{mo1998}.  The latter
represents a likely formation epoch of the Galactic thick disc as
suggested by the age of its stars \citep{edvardsson1993}; (ii) two
stellar morphologies for the satellite: spherical or discy; (iii) two
total (and stellar) mass ratios between the infalling satellite and
the host galaxy (10\% and 20\%); and (iv) three initial orbital
inclinations of the satellites, in both prograde and retrograde
directions with respect to the rotation of the host disc.

Both the host galaxy and the satellite are modelled self-consistently
with both star and dark matter (DM) particles.  The stellar component
of the satellite has structure and kinematics that are consistent with
the observed fundamental plane of dE+dSphs galaxies.  The satellite is
launched far away from the host disc (35--50 times the host disc
scale-length, depending on the experiment) and has orbital parameters
that are consistent with cosmological studies of infalling
substructure \citep{benson2005}.  We will use here the simulations of
20\% mass ratio between the satellite and the host galaxy. This
results in a sample of 24 simulated thick discs. The stellar
components of the satellite and of the host disc are modelled with
10$^5$ particles, implying that the satellite's stars are
over-represented in number (by a factor of five) in the merger
remnant. 

As a reference, Table \ref{summary-thick-disks-paperI} summarises both
the structural and the kinematical properties of the final thick discs
of \citetalias{villalobos-helmi2008}. The kinematical properties
listed here were measured at R=2.4$R_{\rm D}$ for both ``$z$=0'' and ``$z$=1''
experiments, corresponding to $\sim$11 kpc and $\sim$5 kpc,
respectively.  Between brackets we also quote these values for the
``$z$=1'' case at R=3.6$R_{\rm D}$, i.e. $\sim$8 kpc. In the rest of this
paper we focus mainly on the ``$z$=1'' experiments, since the final
stellar mass of the remnant system, $\sim 1.4 \times 10^{10} M_\odot$,
is comparable to that estimated for the Galactic thick disc
\citep[e.g.][]{robin2007}.

For completeness, here we briefly remind the reader of the properties
of the Milky Way thick disc. The measured scalelength of the Galactic
thick disc is comparable to that of the thin disc, i.e. in the range
2.8--4.5 kpc; and its exponential scale-height $z_{\rm 0}$=700--2000 pc
\citep{larsen2003,juric2008}.  Kinematically, the velocity ellipsoid
of the thick disc in the solar neighbourhood is observed to be
($\sigma_{\rm R}$,$\sigma_{\phi}$,$\sigma_{\rm z}$) $\sim$(65,54,38) km s$^{-1}$
\citep{chiba2001,vallenari2006,veltz2008}.  Thick disc stars have a
rotational lag of $\sim$30--50 km s$^{-1}$
\citep{chiba2000,veltz2008}.

A close look at Table \ref{summary-thick-disks-paperI} shows that none
of our simulated thick discs reproduce exactly the properties of the
thick disc of the Milky Way.  An additional point to bear in mind is
that the structure of the merger remnants is likely to change if a new
thin disc is formed from freshly accreted gas, as generally envisioned
in the models explored here \citep[e.g.][]{freeman2002}. However, even if the
detailed properties differ, we believe that studies of the dynamical
phase-space structure of our remnants should give us insight into what
observables can be used for example, to distinguish in-situ stars from
those that have been accreted.

\begin{table} \scalefont{1.0}
\tabcolsep 7.8pt
%\begin{minipage}{126mm}
 \caption{Properties of final thick discs produced in \citetalias{villalobos-helmi2008}.}
 \label{summary-thick-disks-paperI}
% \begin{tabular}{@{}llllllll}
  \begin{tabular}{@{}cccc}
  \hline
``$z$=0''                & 0$\degr$ & 30$\degr$ & 60$\degr$ \\
  \hline
$R_{\rm D}$                  & 4.47     & 4.59      & 3.96      \\
$z_{\rm 0}$                  & 1.25     & 1.63      & 1.78      \\
$\sigma_{\rm R}$             & 94.7     & 84.5      & 69.5      \\
$\sigma_{\phi}$        & 74.3     & 62.8      & 55.4      \\
$\sigma_{\rm z}$             & 46.3     & 53.4      & 54.6      \\
$\overline{\vphi}$  & 119.6    & 122.1     & 150.3     \\
\hline
``$z$=1''                & 0$\degr$ & 30$\degr$ & 60$\degr$ \\
\hline
$R_{\rm D}$                  & 2.26      & 2.28       & 2.04      \\
$z_{\rm 0}$                  & 0.64      & 0.82       & 0.85      \\
$\sigma_{\rm R}$             & 85.7(64.8) & 76.4(53.9)  & 55.4(47.8)      \\
$\sigma_{\phi}$        & 56.7(54.8) & 53.9(44.8)  & 48.2(28.3)      \\
$\sigma_{\rm z}$             & 36.8(20.9) & 41.7(32.5)  & 41.8(42.6)      \\
$\overline{\vphi}$  & 96.1(117.3)& 107.8(120.7)& 122.7(133.7)     \\
  \hline
  \hline
 \end{tabular}
% \medskip
\\
- Scale-lengths ($R_{\rm D}$) and scale-heights ($z_{\rm 0}$) in kpc. Radial 
($\sigma_{\rm R}$), azimuthal ($\sigma_{\phi}$) vertical ($\sigma_{\rm z}$) velocity 
dispersions and mean rotation ($\overline{\vphi}$) in km s$^{-1}$.\\
- Kinematics are measured at 2.4$R_{\rm D}$ for ``$z$=0'' and at 2.4$R_{\rm D}$ (3.6$R_{\rm D}$) 
for ``$z$=1'' experiments.\\
- Only experiments with prograde spherical satellites with masses 20\% of 
$M_{\rm disc,host}$ are listed.
%\end{minipage}
\label{table-tk-paper1}
\end{table}

\subsection{Definition of local volumes}
\begin{figure}
\begin{center}
\includegraphics[width=87mm]{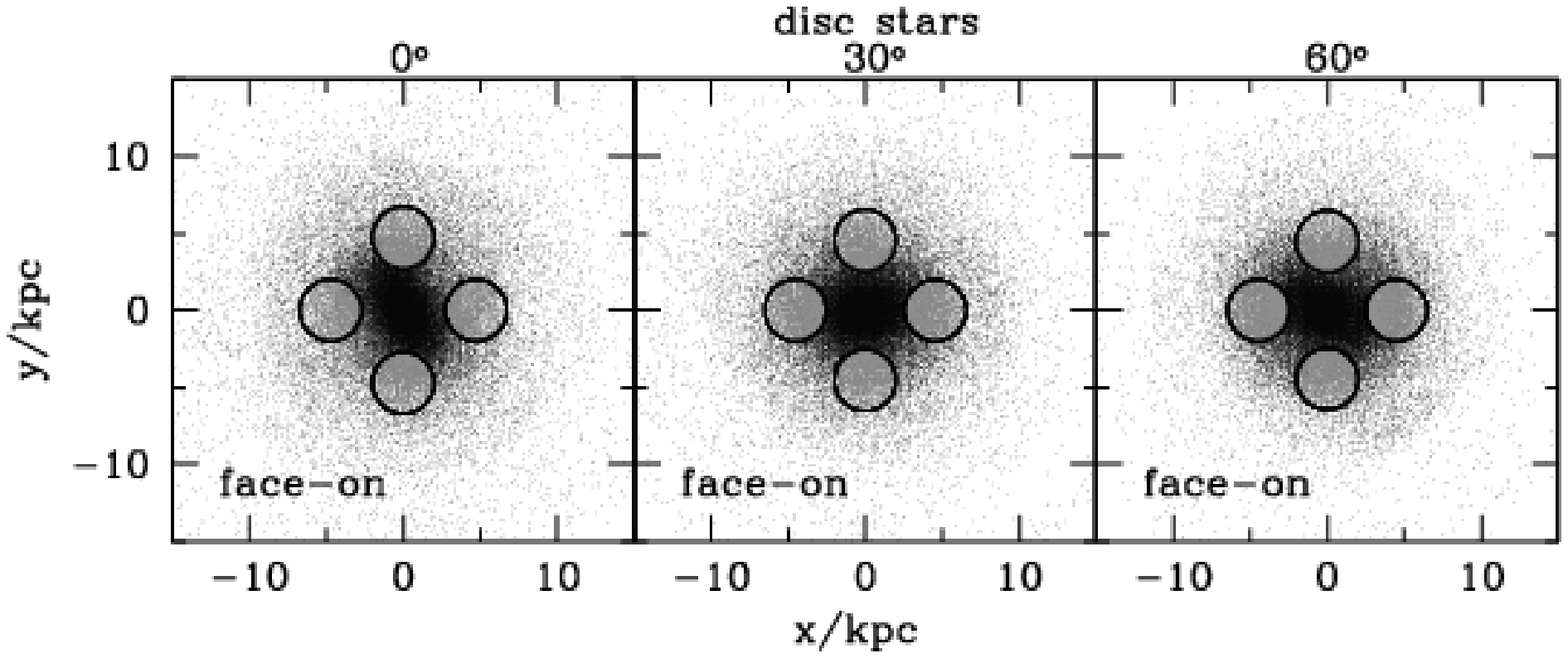}\\
\includegraphics[width=87mm]{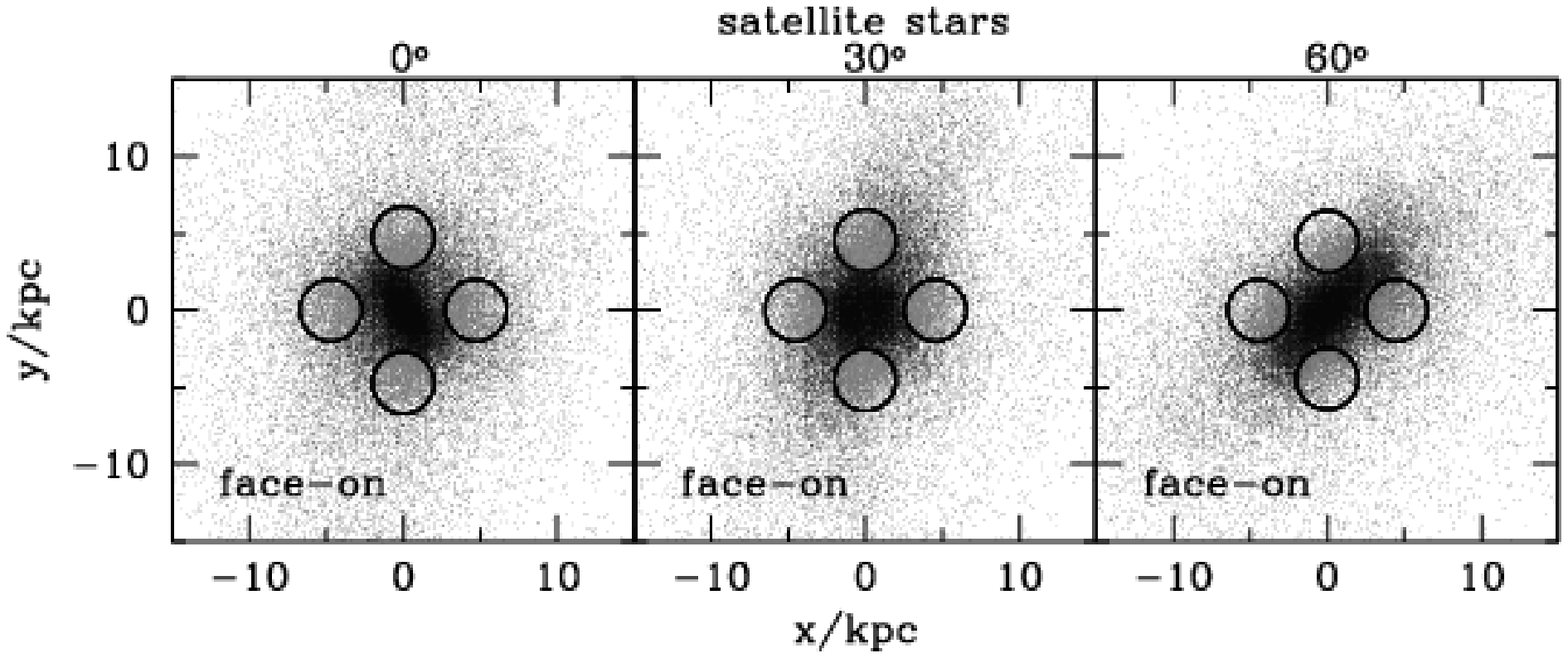}
\caption{Coverage of the spherical volumes defined to analyse the
phase-space structure of the simulated thick discs. The figures show
the final spatial distribution (in a face-on view) of ``$z$=1''
experiments with spherical satellites on prograde orbits for three
different initial orbital inclinations. Volumes are centred at
$R_{\rm v}$=2.4$R_{\rm D}$ ($\sim$5 kpc) from the galactic centre and are 2 kpc
radius. Stars belonging to the heated disc (above) and to the
satellites (below) are plotted separately to highlight the different
spatial distributions.}
\label{coverage-spheres}
\end{center}
\end{figure}

Our interest lies in determining the structure of phase-space of our
merger remnants especially in small volumes which may resemble the
Solar neighbourhood.  This is motivated by our aim to eventually
compare our predictions to observations, and because typically we only
have (access to) full phase-space information for relatively nearby
stars.

Given that the scalelengths of our simulated thick discs are $\sim
35$\% smaller than what has been estimated for the Milky Way, it is not
straightforward to decide which radius would correspond to the ``Solar
circle''. This is why we explore the phase-space structure in volumes
located at two different radii, namely 2.4$R_{\rm D}$ ($\sim$5 kpc) and
3.6$R_{\rm D}$ ($\sim$8 kpc), where $R_{\rm D}$ is the final thick-disc
scalelength in our experiments (see Table \ref{table-tk-paper1}).

\begin{figure*}
\begin{center}
\includegraphics[width=88mm]{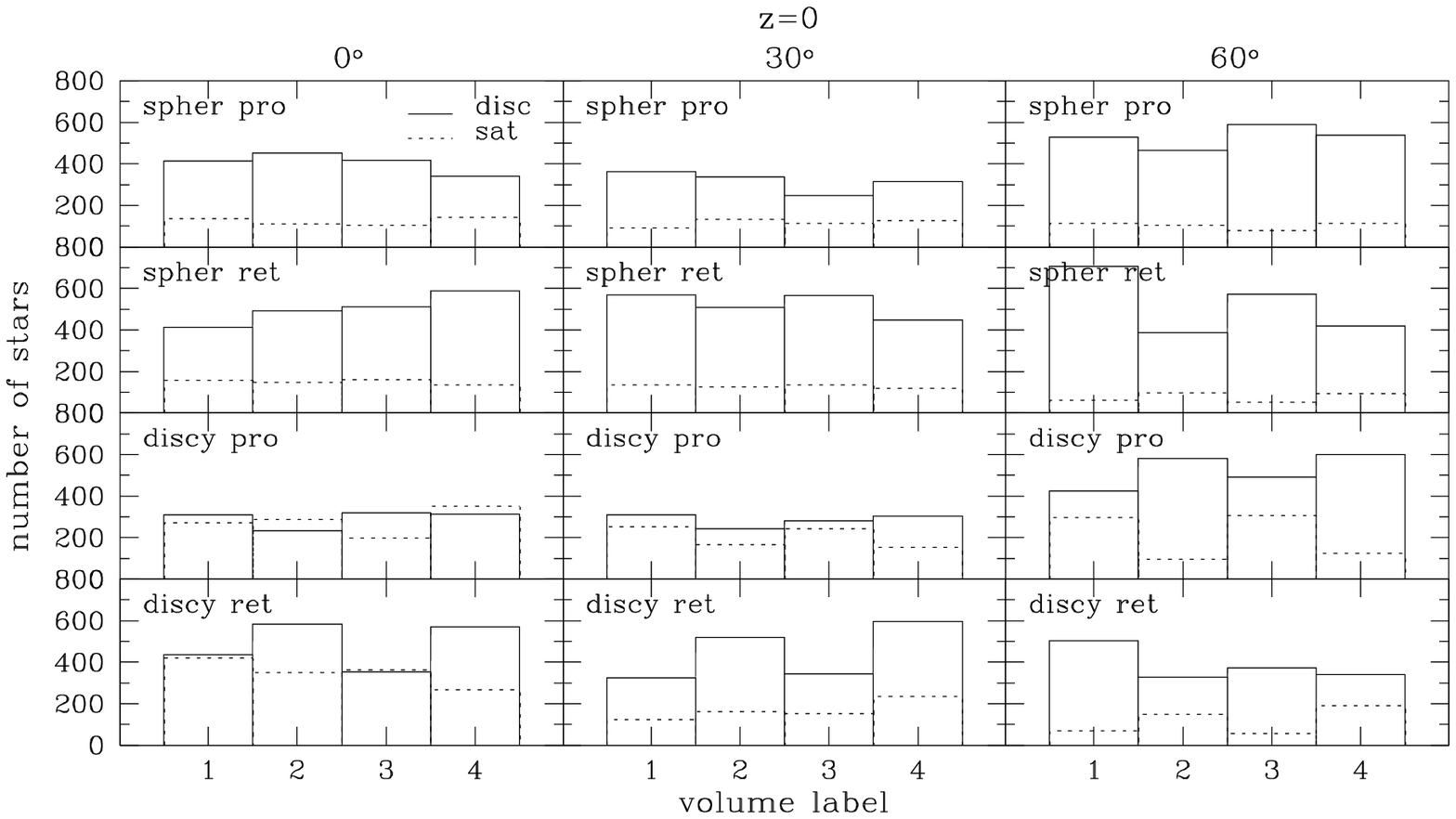}
\includegraphics[width=88mm]{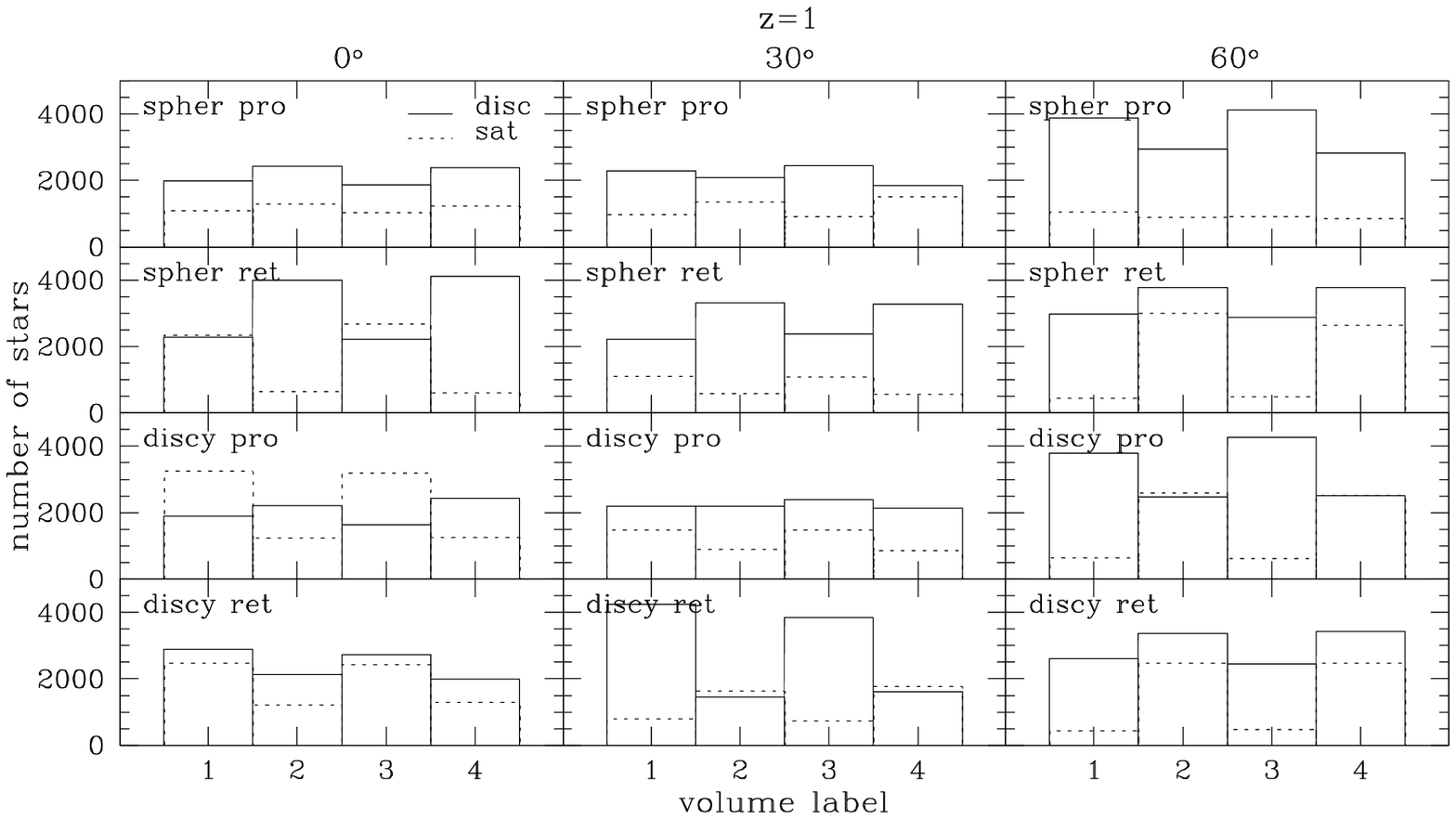}
\caption{ Number of disc and satellite stars within each volume for
all the experiments at ``$z$=0'' and ``$z$=1''.  The volumes are labelled
from 1 to 4 according to their location on the \emph{xy} plane
($+x$,$+y$,$-x$,$-y$, see Fig. \ref{coverage-spheres}).  In general,
the spatial distribution of stars is not uniform for all 4 volumes,
indicating different degrees of deviation from axisymmetry in both the
heated disc and the satellite debris.  }
\label{numpart-per-vol}
\end{center}
\end{figure*}

To obtain phase-space information from particle samples inside such
volumes for our simulated thick disc, we first define a Cartesian
system with the origin in the centre of mass of the thick disc and the
$z$-axis aligned with its rotational axis. The angular momentum vector
points towards the galactic south pole ($-z$), hence the rotation is
clockwise.  We place four identical spherical volumes onto the disc
plane defined as being perpendicular to the angular momentum vector.
The centres of these spheres have coordinates ($+R_{\rm v}$,0,0),
($-R_{\rm v}$,0,0), (0,$+R_{\rm v}$,0) and (0,$-R_{\rm v}$,0). $R_{\rm v}$ is defined as
either 2.4$R_{\rm D}$ or 3.6$R_{\rm D}$.  We explore spheres of several sizes but,
unless noted otherwise, the rest of the paper will refer to volumes of
2 kpc radius.

Fig. \ref{coverage-spheres} illustrates the volumes' coverage of the
final thick discs for the case of the ``$z$=1'' experiment using a
prograde spherical satellite with initial inclination of
30$\degr$. The volumes are located at $R_{\rm v}$=2.4$R_{\rm D}$ from the galactic
centre and are labelled from 1 to 4 according to where they are placed,
i.e., ($+x$,$+y$,$-x$,$-y$)$\equiv$(1,2,3,4).

In the next Section we will characterise the velocity distribution of
stars in these local volumes. Because our remnant discs are not fully
axisymmetric as evidenced in Fig. \ref{coverage-spheres}, it is
important to establish first how the properties of the velocity
distribution function depend on location. Therefore in Section
\ref{sec:bars} we quantify the deviations from axisymmetry and
establish their impact on the velocity distribution. We then focus on
the dynamical properties of the in-situ stars versus those accreted in
Section \ref{sec:in-situ.vs.accreted} and Section \ref{sec:stats}.

\section{Results}
\label{sec:results}

\subsection{Effect of non-axisymmetries}
\label{sec:bars}

\subsubsection{On the spatial distribution}
\label{nonaxis-spatial}

Fig. \ref{coverage-spheres} shows that the distributions of heated
disc particles and satellite particles are not symmetric with respect
to the rotation axis of the system.  Note that there does not seem to
be a correlation in the deviations from axisymmetry between the disc 
and the satellite stars. This suggests that these asymmetries are likely 
to have different origin.
  
Fig. \ref{numpart-per-vol} gives a general overview, for all 24 of our
experiments, of the distributions of stars in the final thick discs.
This figure shows the variation of the number of stars from volume to
volume for each experiment. Note that, in order to facilitate the
comparison, the number of satellite stars has not been normalised
according to mass ratio between host disc and the stellar component of
the satellite (see \citetalias{villalobos-helmi2008}), meaning that they are over-represented by
a factor of five. Overall, in each experiment, both disc and satellite
present some degree of volume-to-volume alternation in the number of
stars they contribute, demonstrating their asymmetric spatial
distributions. In general, this alternation is different for the disc
and for the satellite stars.

\begin{figure}
\begin{center}
\includegraphics[width=88mm]{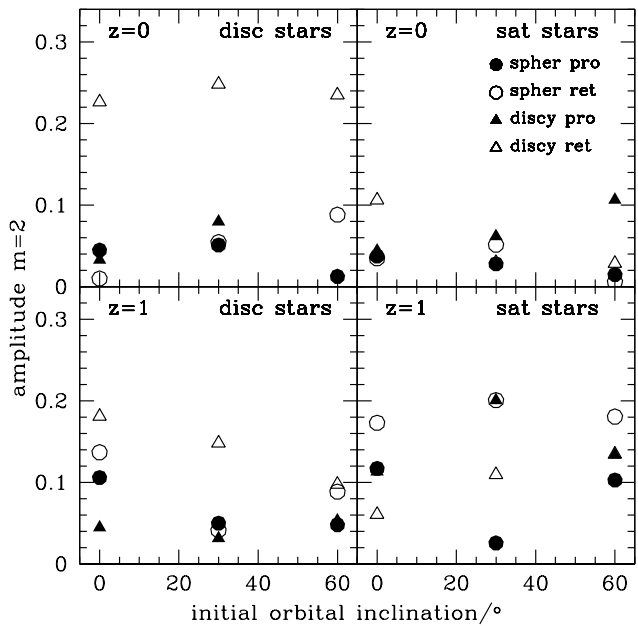}
\caption{Global amplitudes of the $m=2$ deviations from axisymmetry in
the final spatial distributions of both disc (left panels) and
satellite (right panels) stars for all experiments at ``$z$=0'' (upper
panels) and ``$z$=1'' (lower panels).}
\label{ampl-bar-vs-morph-incl}
\end{center}
\end{figure}

\begin{figure*}
\begin{center}
\includegraphics[width=55mm]{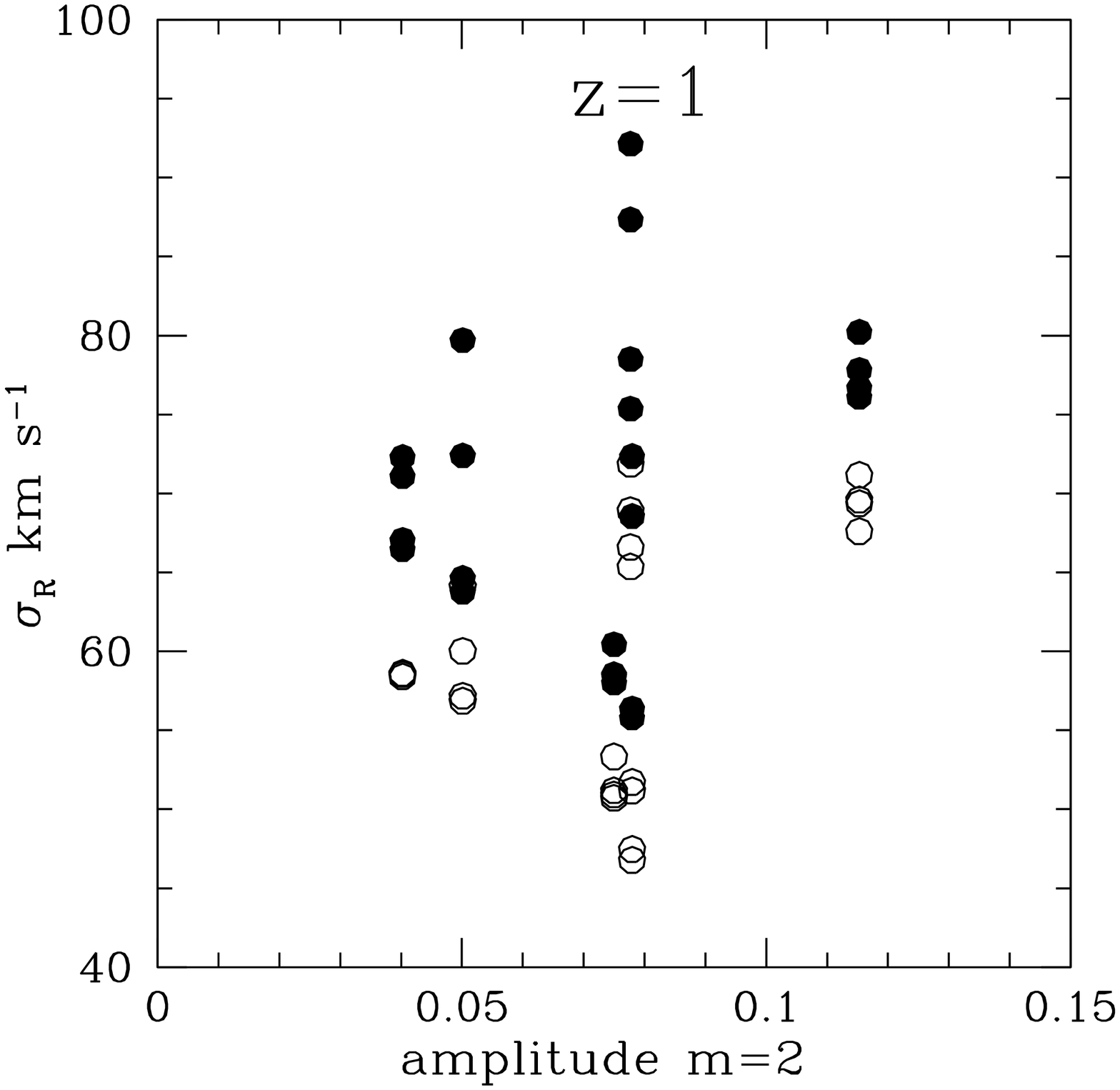}
\includegraphics[width=55mm]{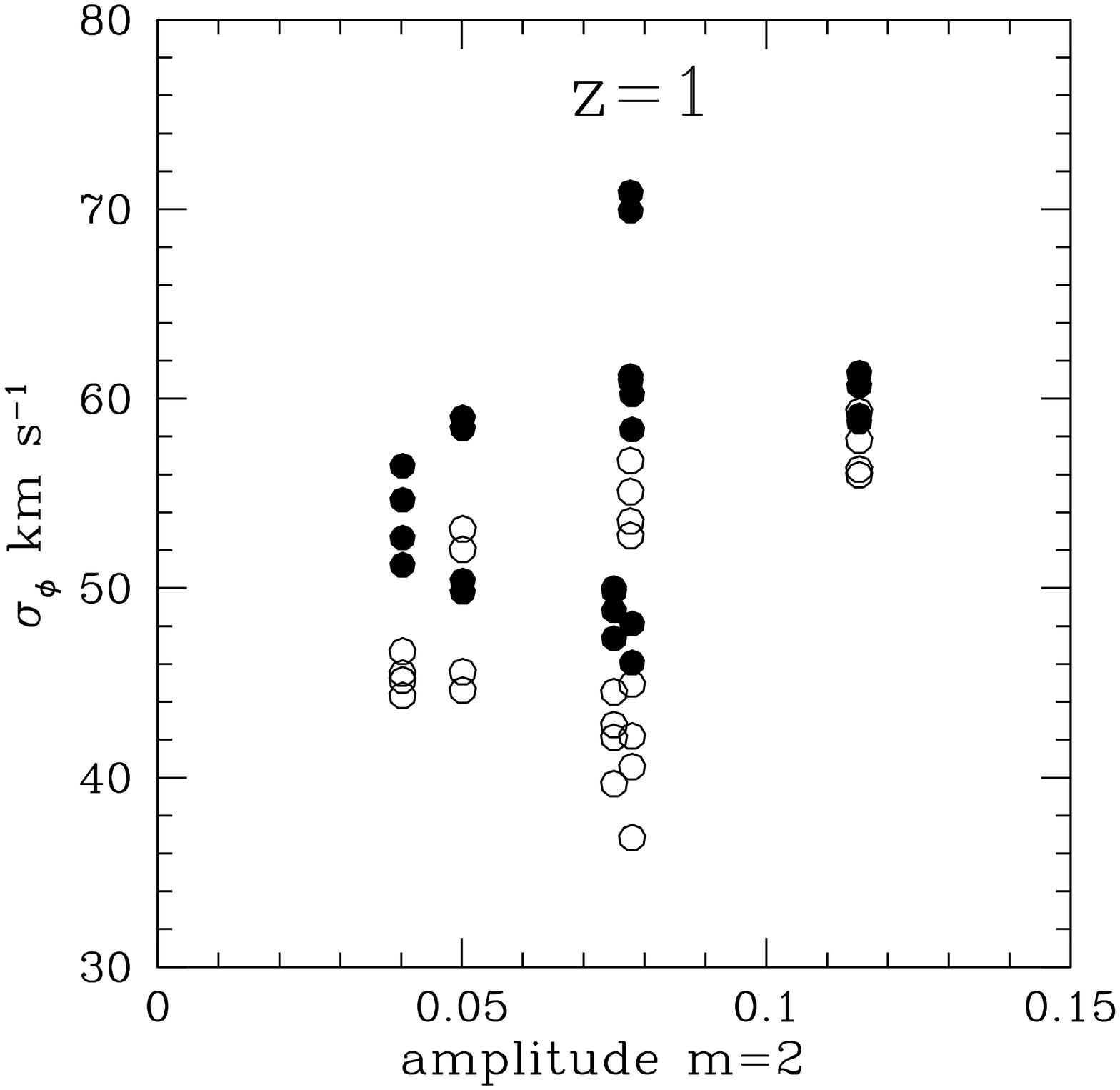}
\includegraphics[width=55mm]{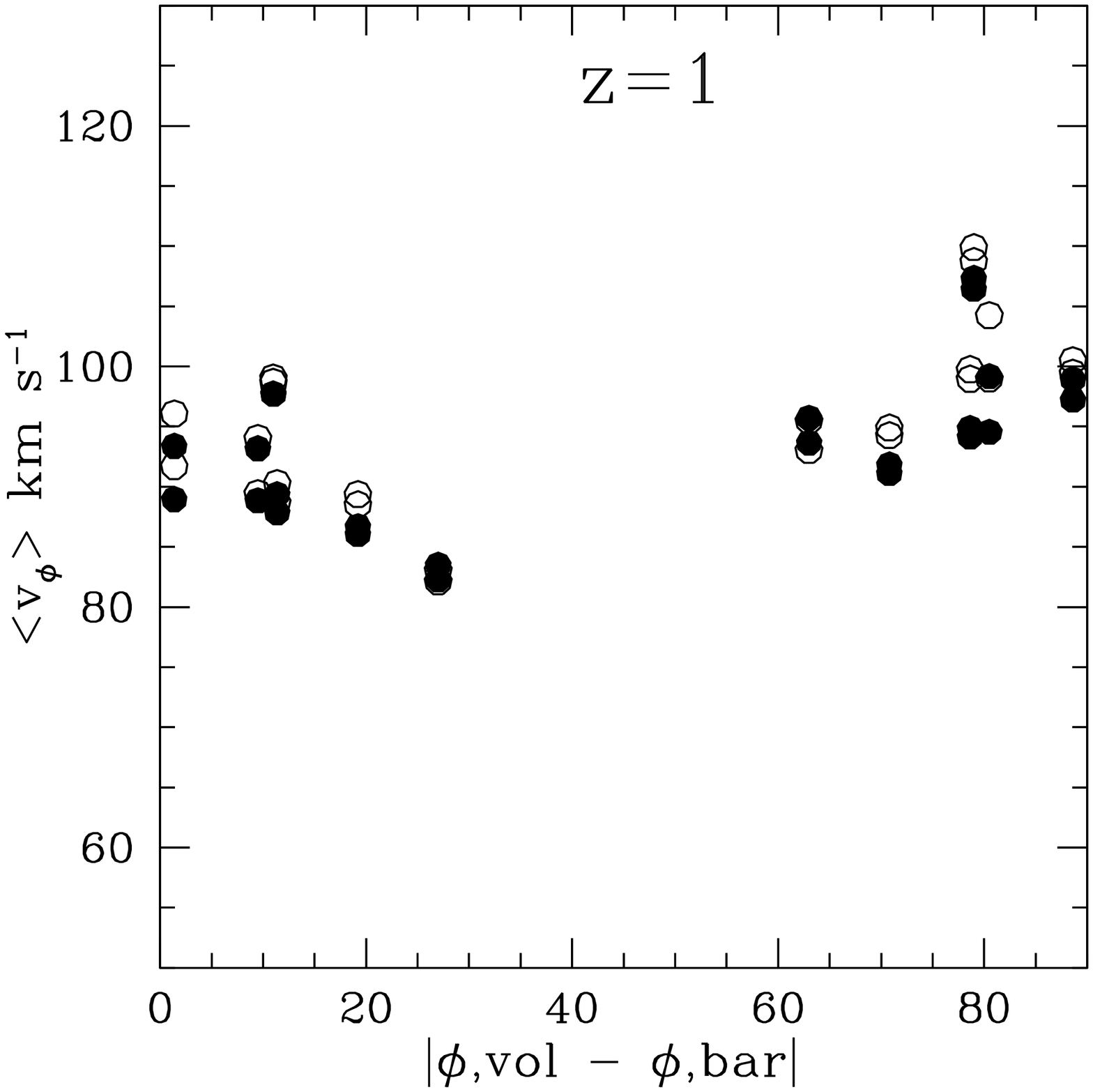}
\caption{Radial (left panel) and tangential (central panel) velocity
dispersions in each volume as a function of the amplitude of the $m=2$
deviations from axisymmetry in the spatial distribution of disc stars
for all prograde ``$z$=1'' experiments.  Velocity dispersions are shown
considering both disc and satellite stars (solid circles) and disc
stars only (open circles). The panel on the right shows the mean
rotational velocity in each volume as a function of the angular
separation between the volume and the major axis of the ``bar'' formed
by the disc stars, for all ``$z$=1'' experiments.}
\label{vrsig-vpsig-vs-bar}
\end{center}
\end{figure*}

Fig. \ref{ampl-bar-vs-morph-incl} shows the global amplitudes of the
$m=2$ deviations of both disc (left panels) and satellite (right
panels) spatial distributions as a function of the initial orbital
inclination of the satellites, for all 24 experiments.  The amplitudes
are measured by first binning the final spatial distributions of stars
in cylindrical shells, within $|z|<5$ kpc and $|z|<3$ kpc for experiments
at ``$z$=0'' and ``$z$=1'', respectively. 
Then for each bin the Fourier components of the
second harmonic of the angular distribution are computed as:
\begin{equation} \label{fourier-comp}
a_2 = \frac{1}{N} \sum^{N}_{i=1} \sin (2\phi_i), \qquad b_2 = \frac{1}{N} \sum^{N}_{i=1} \cos (2\phi_i),
\end{equation}
(\citealt{valenzuela2003} and references therein) with $N$ the number
of stars in each bin and $\phi_i$ the stars' angular position. For
each bin the amplitude of the second harmonic is
$A_2^2=(a_2^2+b_2^2)/2$.  The global amplitude is then defined as
$\langle A_2 \rangle = \sqrt{\langle A_2^2 \rangle}$, averaging over
radii $R<1.5R_{\rm D}$. Additionally, the angular phase of the $m=2$
deviations are measured by finding the principal axes of the inertia
tensors of the spatial distribution of stars projected onto the $xy$
plane. In general, strong disc asymmetries are present out to a
distance of $\sim$3--4 kpc ($\sim 1.7R_{\rm D}$) for the ``$z$=1''
experiments.

Figure \ref{ampl-bar-vs-morph-incl} shows that in general, for both
disc and satellite stars, there are no clear dependencies of the $m=2$
amplitude with inclination. However it is interesting to notice that
discy satellites on retrograde orbits have induced the largest $m=2$
deviations on the heated discs.  This can be explained by the fact
that this type of satellite loses mass faster as it evolves in
comparison to spherical satellites (see Fig. 3 in \citetalias{villalobos-helmi2008}). This
results in a smaller drag force and hence in a longer decay
timescale. Moreover, in general satellites on retrograde orbits are
found to need more time to decay (see Fig. 2 in \citetalias{villalobos-helmi2008}) which is a
consequence of the weaker dynamical friction exerted by the main disc
as compared to the prograde orbit case \citep[see
also][]{velazquez1999}. Therefore, the aforementioned satellite
configuration induced a longer lasting and more powerful asymmetric
perturbation on the main disc, eventually leading to the formation of
a bar with a larger amplitude.

On the other hand, the $m=2$ deviations for the satellite debris seem
to reach similar amplitude independently of the satellites'
morphologies and initial orbital inclinations. This suggests that
$m=2$ deviations for disc and satellite stars may be caused by
different dynamical effects.  The satellites in our simulations have
rather radial orbits (apo-to-peri ratios $\sim 5$, see also Section 3.1 in
\citetalias{villalobos-helmi2008}), for all initial orbital inclinations. This implies that
their angular momenta are relatively low, or in other words that the
amount of energy associated to tangential motions is much smaller than
that associated to motion in the radial direction. This in turn, means
that the precession rate of these orbits is slow. Furthermore, our
satellites are relatively cold in comparison to the host galaxy. These
factors make the range of orbital angular frequencies of satellite
stars quite small, implying that that the mixing in the angular
direction has to take place on fairly long timescales. Thus as the
satellite gets disrupted, its stars remain somewhat constrained in
azimuth, which results in the observed $m=2$ deviations, in rather
close analogy to what happens as a result of the radial orbit
instability \citep{bt2008}.

\begin{figure*}
\begin{center}
\includegraphics[width=35mm]{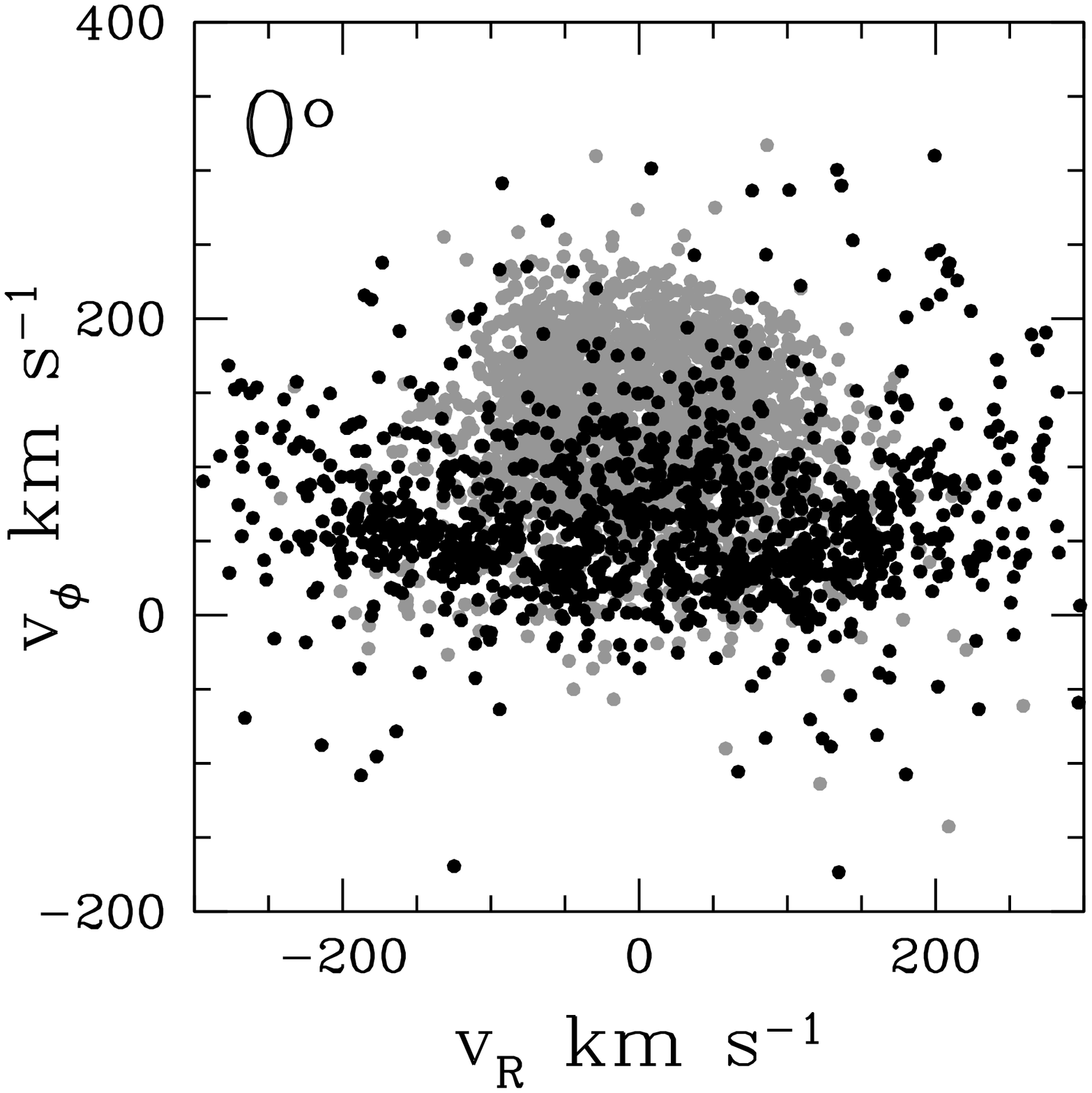}
\includegraphics[width=35mm]{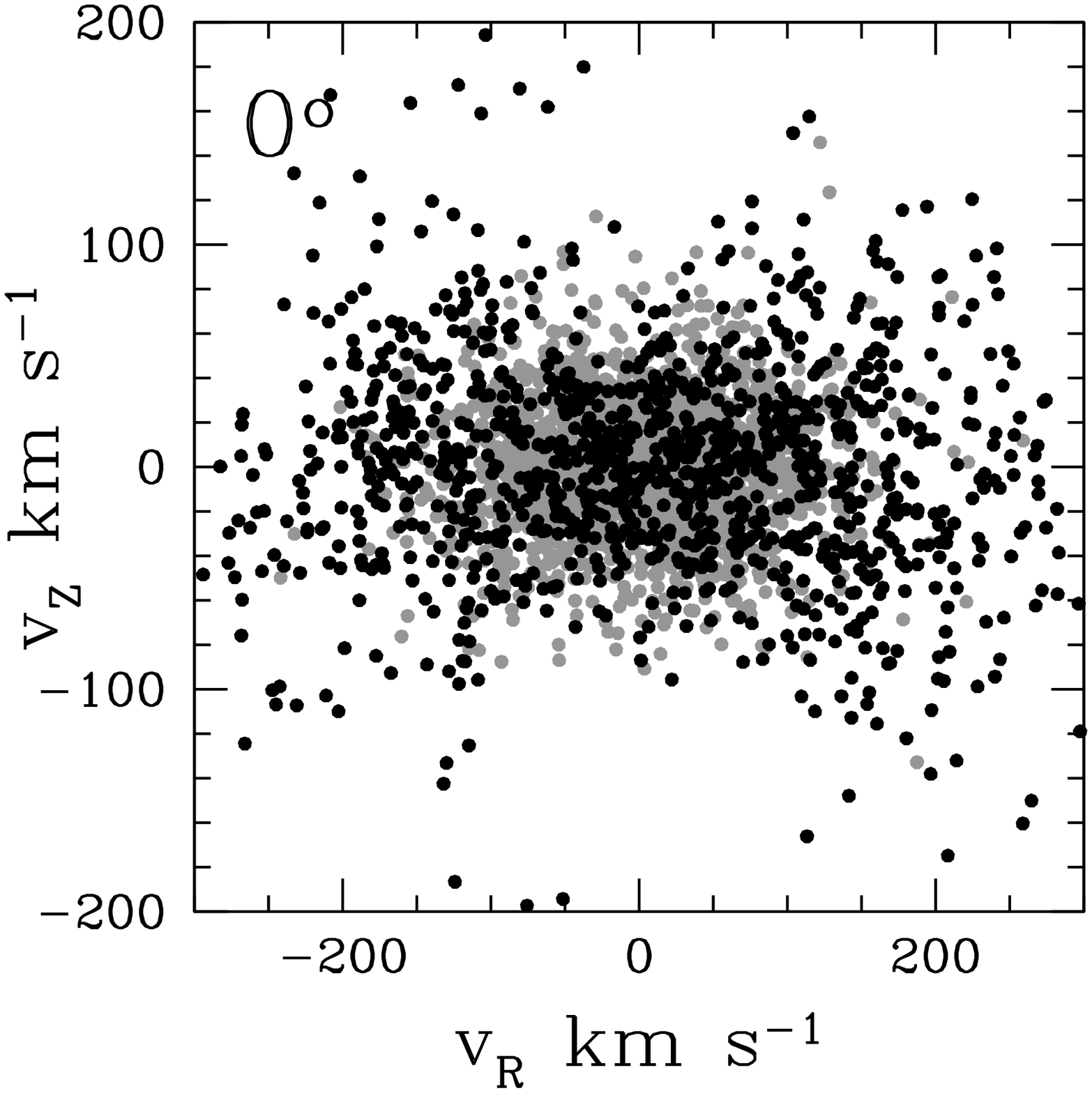}
\includegraphics[width=35mm]{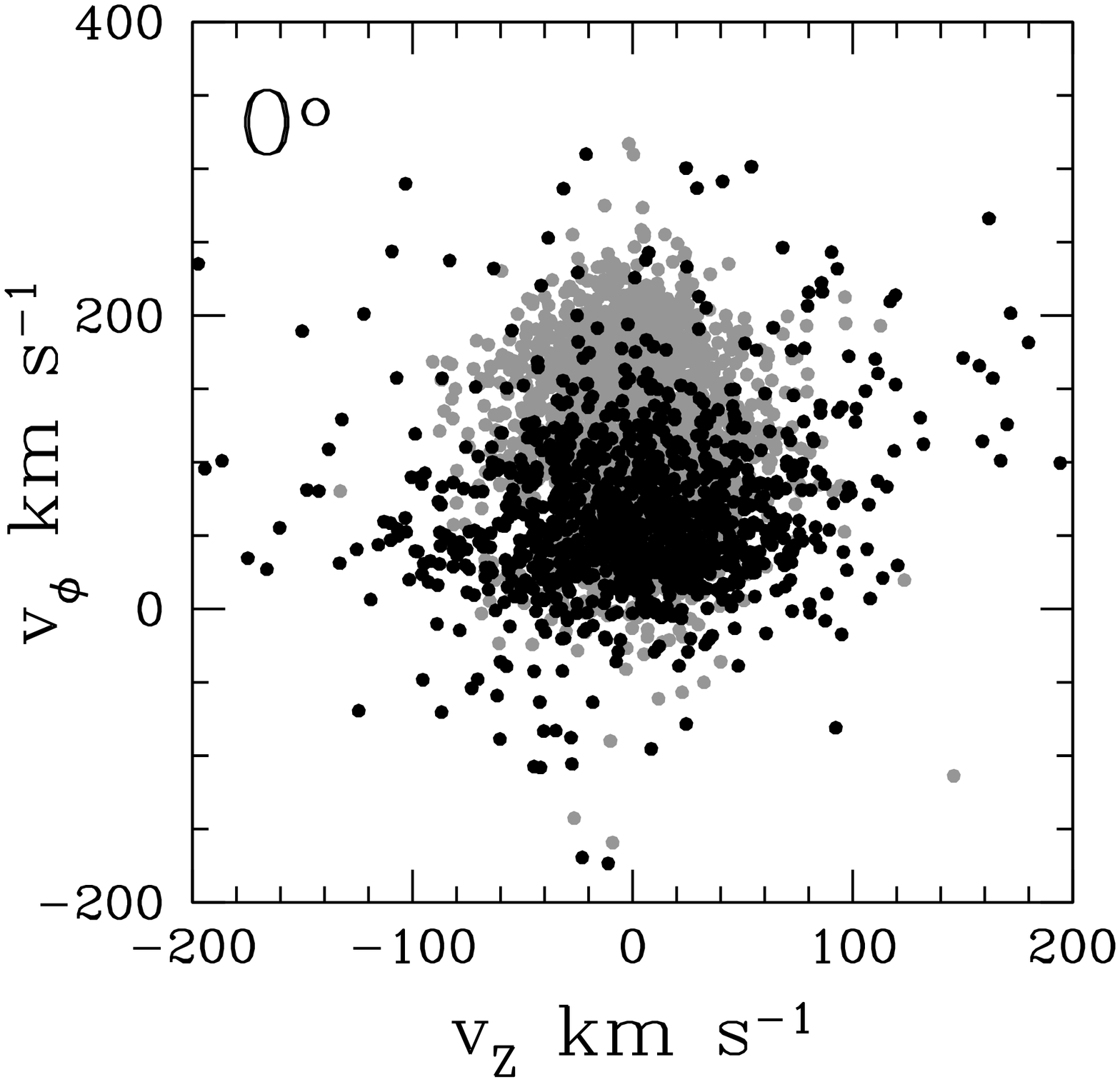}\\
\includegraphics[width=35mm]{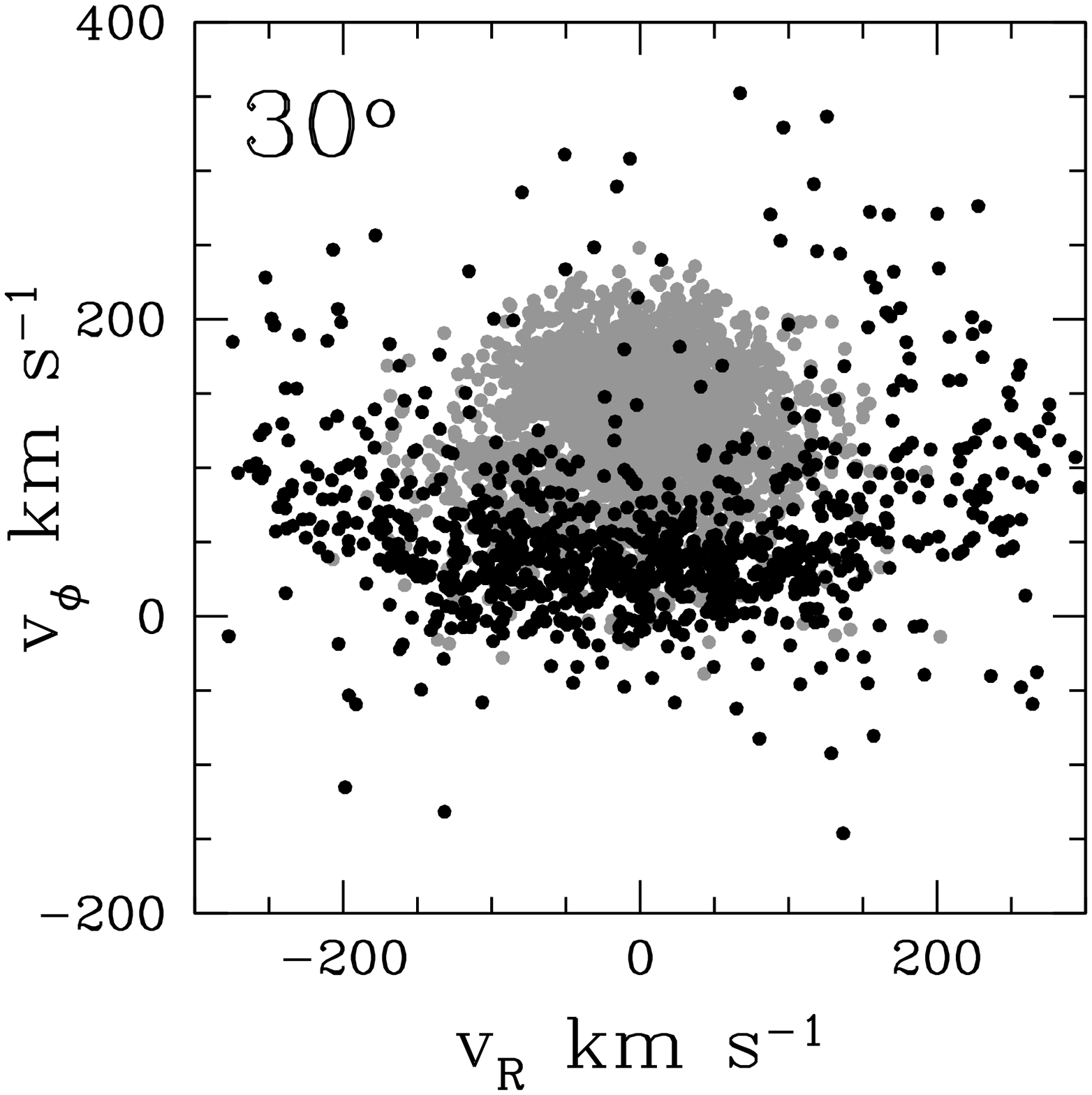}
\includegraphics[width=35mm]{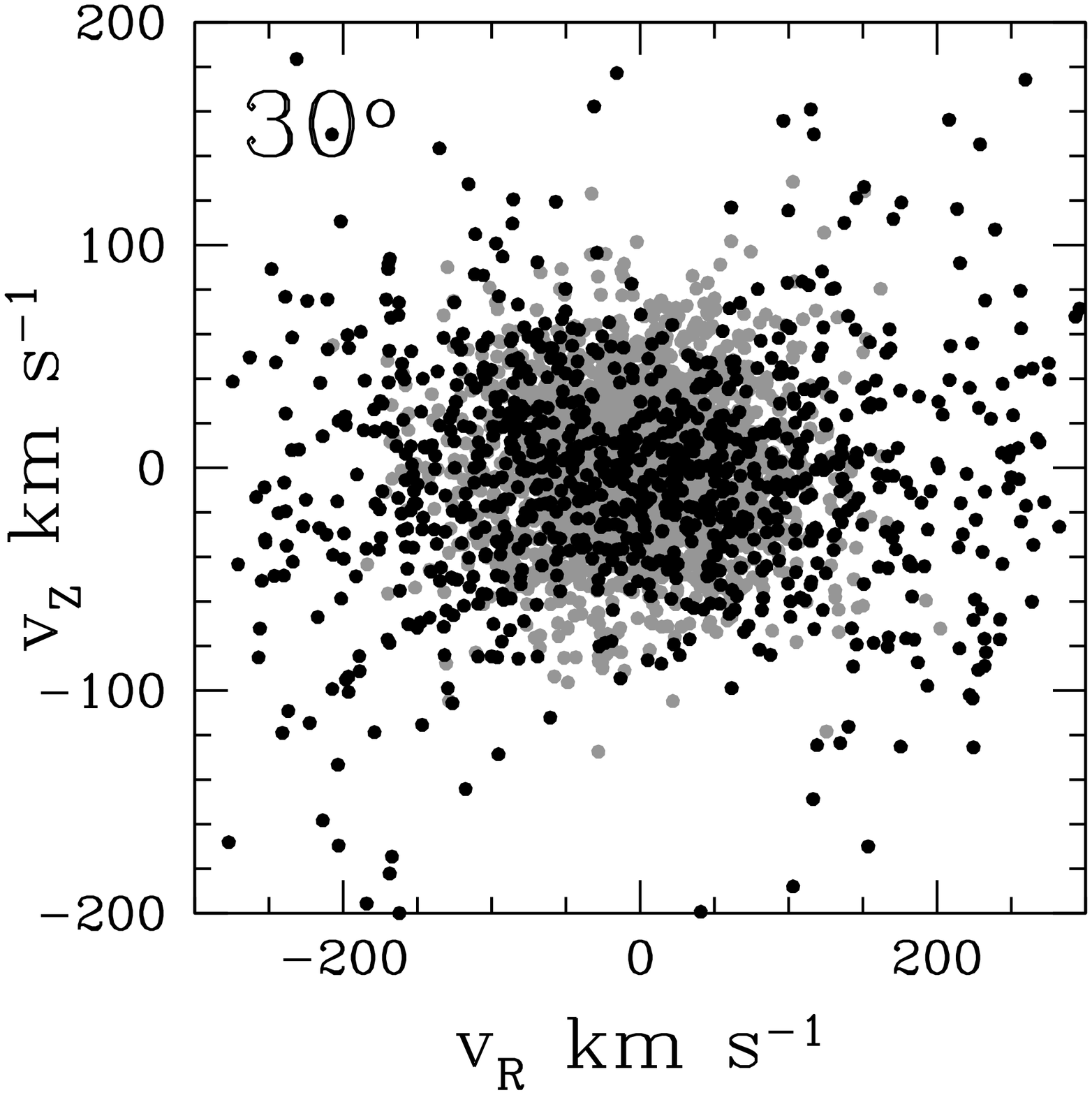}
\includegraphics[width=35mm]{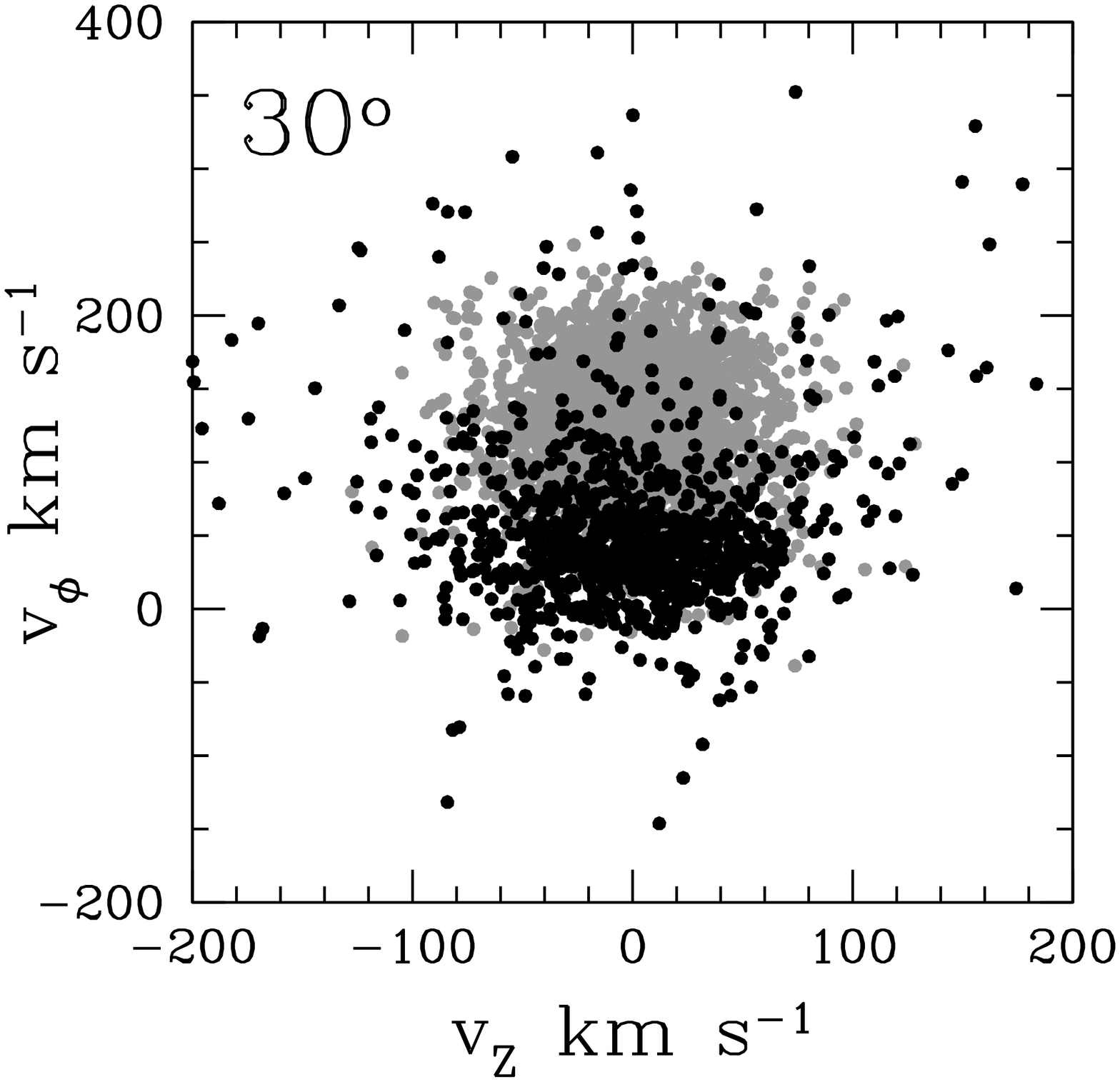}\\
\includegraphics[width=35mm]{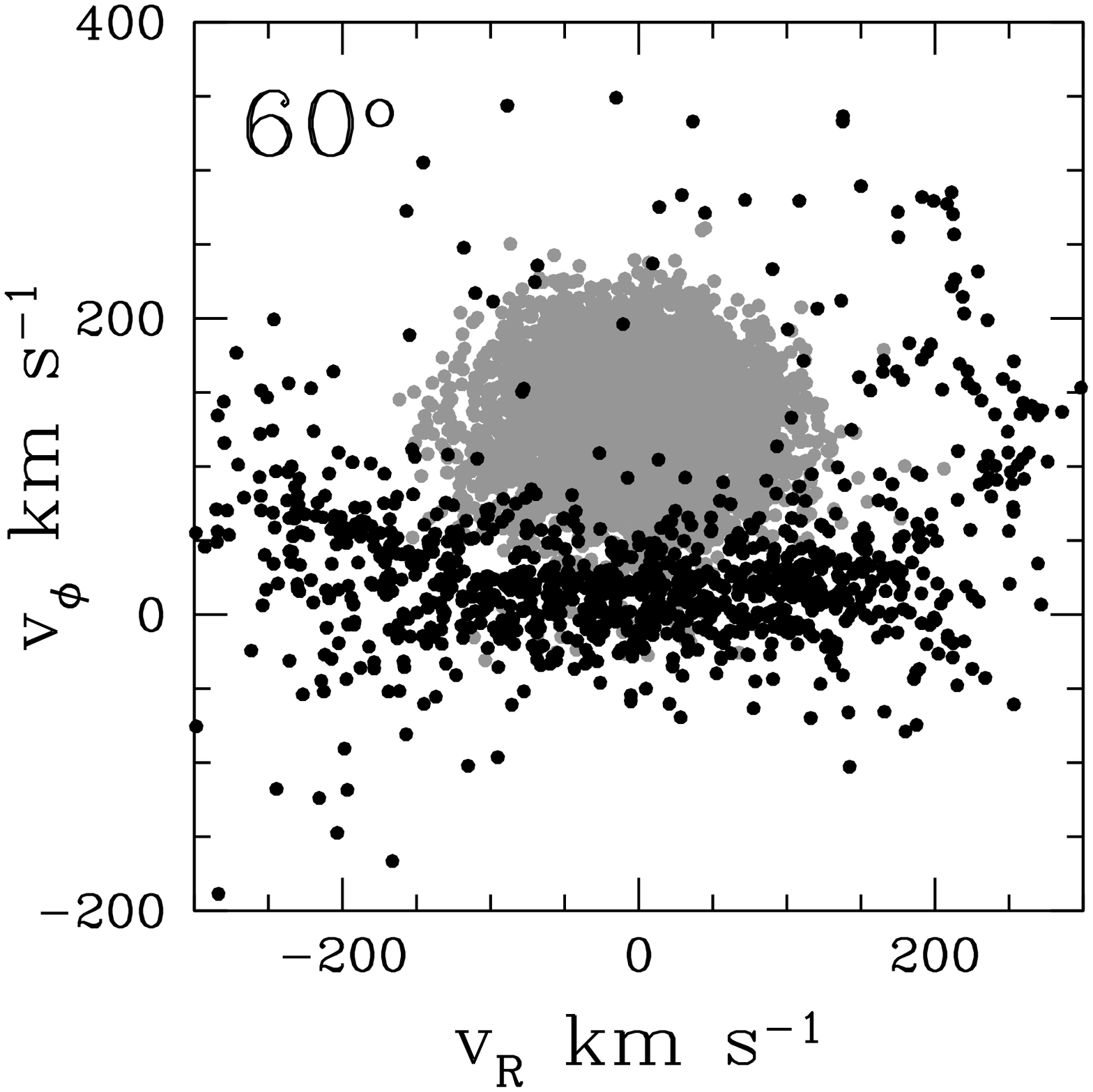}
\includegraphics[width=35mm]{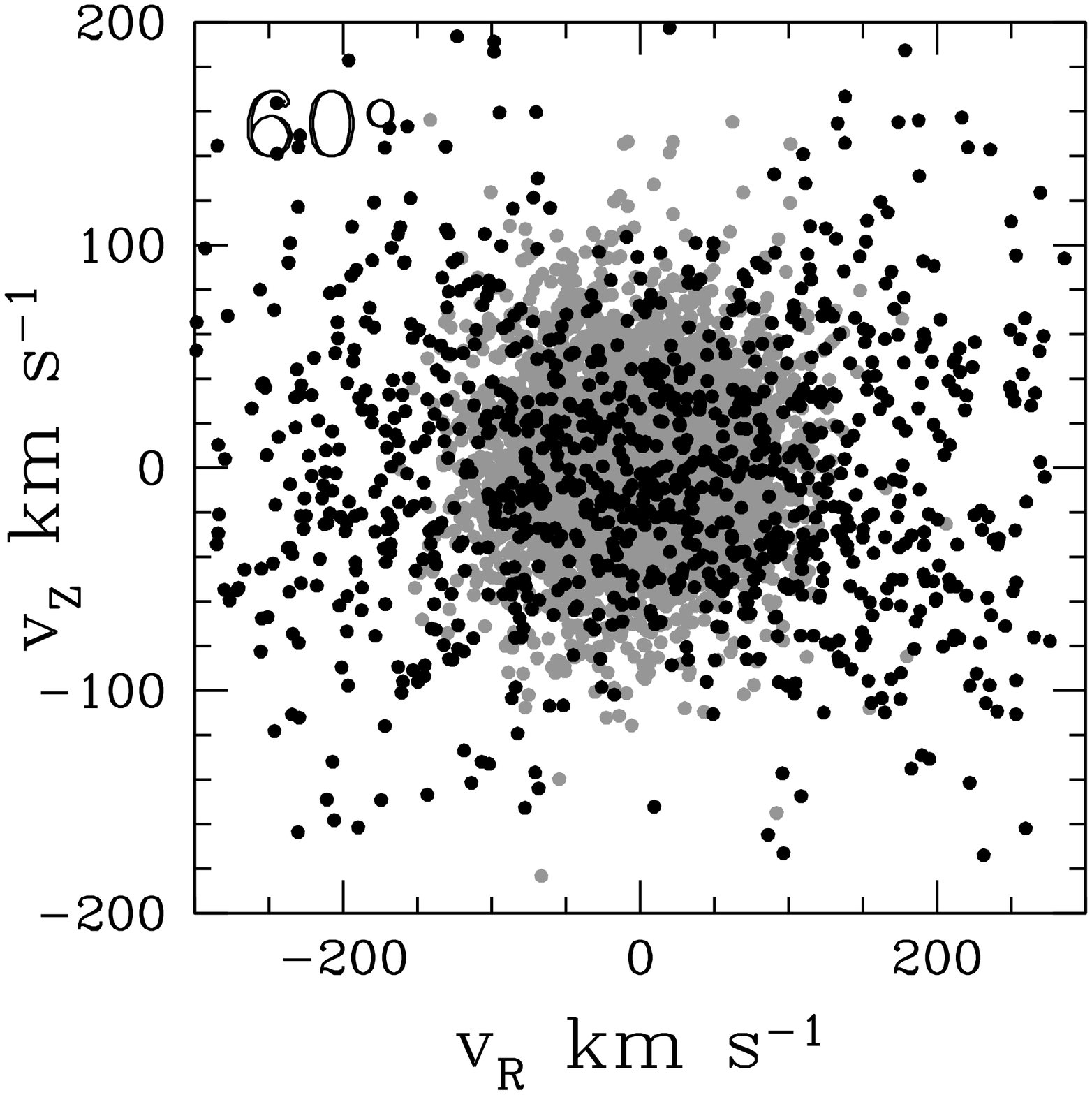}
\includegraphics[width=35mm]{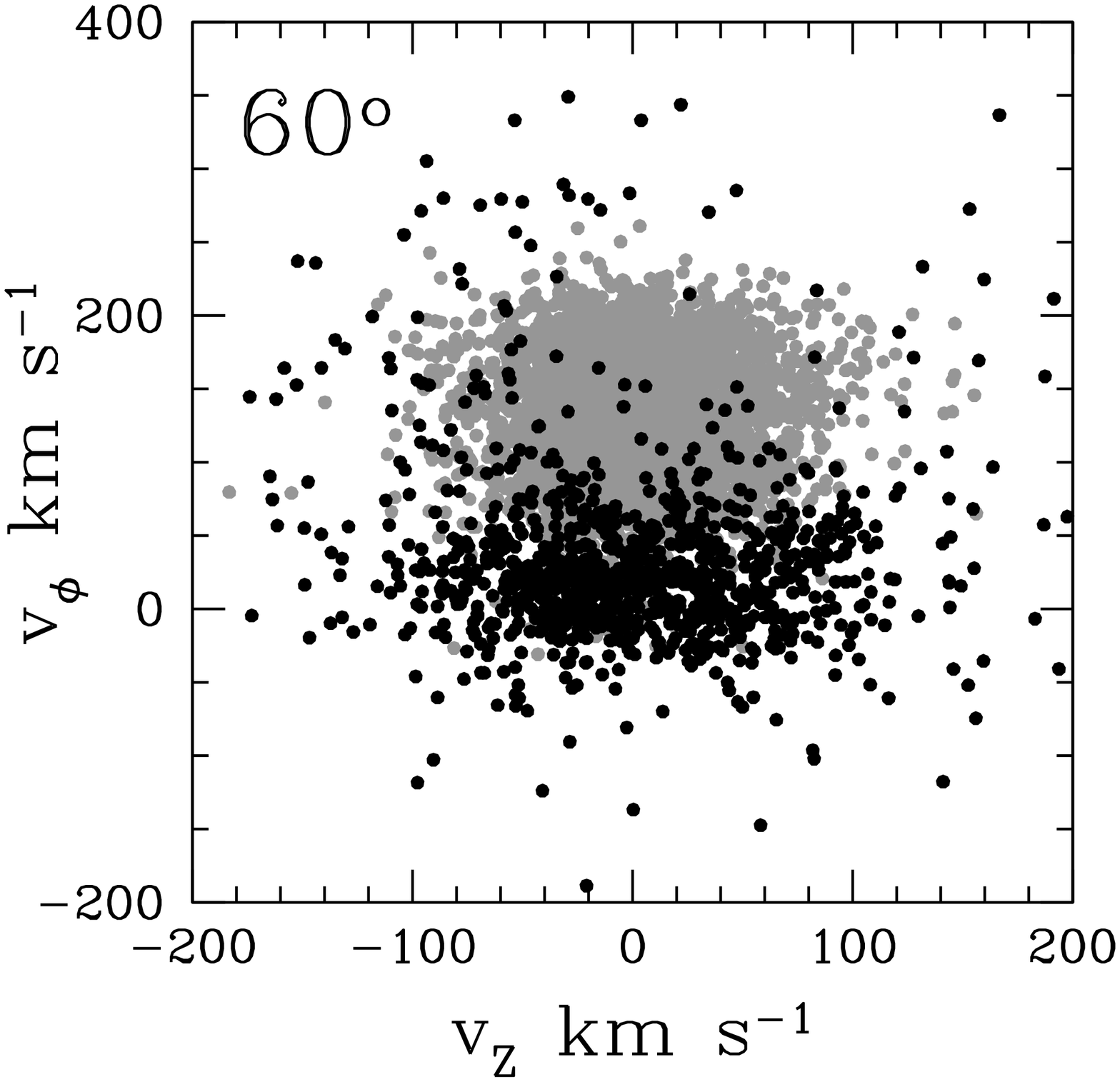}
\caption{ Final velocity distribution of disc (grey) and satellite
(black) stars for different initial orbital inclinations of a prograde
spherical satellite: 0\degr(upper row), 30\degr (middle) and 60\degr
(bottom); corresponding to ``$z$=1'' experiments.  Stars are within
spheres of 2 kpc radius located at R=2.4$R_{\rm D}$ from the galactic
centre. All satellite stellar particles have been plotted here,
implying that they are over-represented by a factor of five. }
\label{vrvpvz}
\end{center}
\end{figure*}

There are two important implications of the characterisations made
above. Firstly thick discs could well be asymmetric because of the
presence of a small central bar. Secondly, the fraction of accreted
stars may well vary with azimuthal angle in the plane of the thick
disc. In recent years, \citet{parker03,parker04} have found evidence
that indeed the Galactic thick disc may be asymmetric. More recently,
the suggestion has been made that the stellar asymmetry of faint
thick-disc/inner-halo stars in the first quadrant ($l = 20\degr-45\degr$)
could be an indication of a triaxial thick disc or a merger
remnant/stream \citep{larsen08}.

\subsubsection{On the kinematics}
\label{on-kine}
In order to investigate the possible effects of the ``bars'' on the
kinematics of stars in our local volumes, we have computed several
statistical moments of the velocity distributions (mean, dispersion,
skewness and kurtosis). We have then correlated these properties to
both the location and the amplitude of the $m=2$ deviations of the
spatial distribution of disc stars (since this bar dominates in terms
of mass/surface density as compared to the triaxial distribution of
accreted stars which never dominate the central regions nor near the
plane, see \citetalias{villalobos-helmi2008}).  In general, we do not find a significant
dependence of the moments with respect to the strength of $m=2$
deviations. This is shown explicitly in the left and central panels of
Fig. \ref{vrsig-vpsig-vs-bar}, where we have plotted the radial and
tangential velocity dispersions within each volume, for all prograde
experiments at ``$z$=1'', as a function of the induced global $m=2$
amplitudes. Note that similar results are obtained when the
dispersions are computed considering either disc stars only (open
circles) or disc and satellite stars jointly (solid circles).  Only
weak trends are found for the mean rotational velocity $\langle v_\phi
\rangle$ when computed as a function of the relative location of the
volume with respect to the position angle of the bar, as shown in the
rightmost panel of Fig. \ref{vrsig-vpsig-vs-bar}. When the various
moments are computed in volumes which are located farther away from
the centre, such as at 3.6$R_{\rm D}$, the correlations are even less
prominent.

\begin{figure}
\begin{center}
\includegraphics[width=89mm]{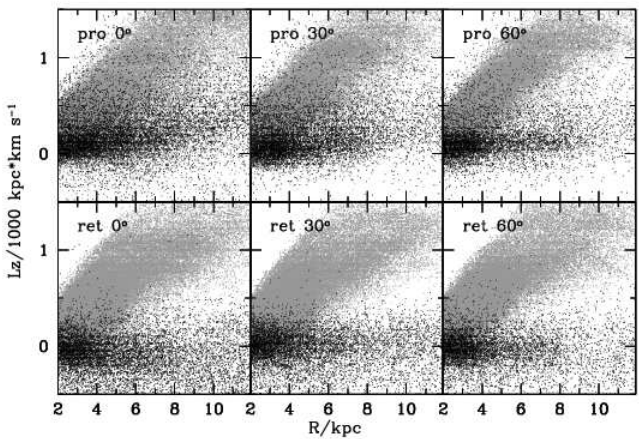}
\caption{ Final distribution of $\Lz$ of disc (grey) and satellite
stars (black; only one in five is shown) as a function of galactic
radius, for prograde (top row) and retrograde (bottom row) satellites
with initial orbital inclinations 0\degr (first column), 30\degr
(second) and 60\degr (third). Stars are located within $2<R<12$ kpc
and $|z|<1$ kpc.  }
\label{grid-lzevol}
\end{center}
\end{figure}

\subsection{The kinematics of disc and satellite stars}
\label{sec:in-situ.vs.accreted}

Fig. \ref{vrvpvz} shows the final velocity distributions of both disc
(grey) and satellite (black) stars for the ``$z$=1'' experiments with a
spherical satellite with initial inclinations of 0$\degr$, 30$\degr$
and 60$\degr$ on a prograde orbit.  The stars plotted in this figure
are enclosed within a spherical volume centred at $R=2.4R_{\rm D}$.

This Figure shows that disc and satellite stars have radically
different distributions in velocity space. For instance, in the
$\vR-\vphi$ plane, satellite stars are distributed in a
banana-shape, whereas the disc stars define a centrally concentrated
clump.  This difference is a consequence of the more eccentric
orbits of the accreted stars compared to those from the disc. The
accreted stars that cross the volume under consideration have slightly
different orbital phases, i.e. stars with $\vR=0$ are at the apocentre
while those travelling either towards (or away from) it have $\vR>0$
(or $\vR<0$), leading to this very characteristic banana-shape
\citep[see also][]{hw99}. In the $\vR-\vz$ plane satellite stars
populate mostly the outskirts of the distributions, with a clear
dependence on the initial orbital inclination. Stars from the
satellites on more inclined orbits show larger vertical velocities as
they cross the disc plane as expected.  Both the $\vR$ and $\vz$
distributions are very symmetric, showing that the stellar particles
are well mixed in these directions by the end of the
simulation. Satellite stars typically have lower mean rotational
velocity in comparison to disc stars, as can be seen in the
$\vz-\vphi$ plane.

In Fig. \ref{grid-lzevol} we show the final distributions of the
$z$-component of the angular momentum ($\Lz$) of both disc and
satellite stars as a function of cylindrical radius, for ``$z$=1''
experiments with a discy satellite.  In this figure, stars are located
within $2<R<12$ kpc and $|z|<1$ kpc.  The trend followed by the disc
stars reflects a constant rotational velocity (hence $\Lz \propto
R$). The stars from the satellite, on the other hand, do not show this
same trend, since their $z$-angular momenta are relatively constant
with radius; the absolute value being related to the initial
conditions of the satellite's orbit. This is why the separation
between both types of stars has a mild trend with initial orbital
inclination of the satellite, especially for prograde orbits, and also
why the retrograde cases have negative $\langle \Lz \rangle$.  This
figure suggests that in order to distinguish more easily between disc
and satellite stars, it is better to study the kinematics at large
galactocentric distances.

Note that these conclusions can be generalised rather easily, and are
quite independent of the particular initial configuration of the
merger.  For example, an increase in the initial tangential velocity
of the satellite, will lead to a higher final value of $\Lz$, implying
that the distinction between accreted and in-situ stars will be more
difficult near the Sun. In such cases, one would have to compile
samples of stars located at much larger distances in the plane, such
that the final $\Lz$ distribution at this location would be bimodal,
as seen for example, at $R \sim 8$ kpc in Figure~\ref{grid-lzevol}. 

It is also important to stress here that the trend of $\Lz$ as
function of $R$ observed for the stars in the heated disc, constitutes
a rather clean test of the formation scenario for the Galactic thick
disc studied here. Such a trend would be expected only in the case a
pre-existing thin disc was present at early times, since only then
most of the stars would be moving on nearly circular orbits.

\begin{figure}
\begin{center}
\includegraphics[width=85mm]{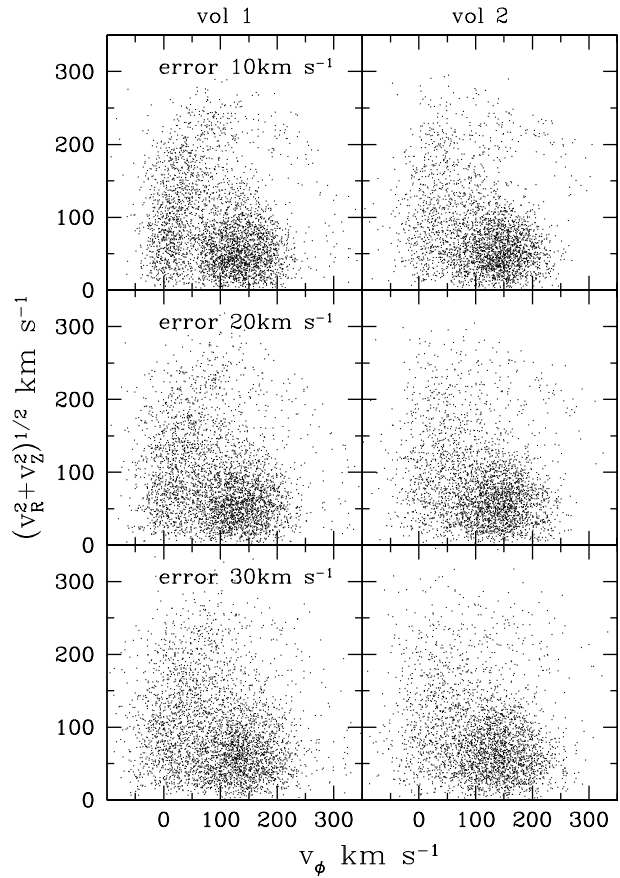}
\caption{ Toomre diagrams of ``$z$=1'' experiments for prograde discy
satellites with initial orbital inclination 30$\degr$ after convolution of
each velocity component with a Gaussian error of: 10 (upper row), 20
(middle) and 30 (bottom) km s$^{-1}$.  In the case of errors $\sim$10
km s$^{-1}$, disc and satellite stars are easily distinguishable (disc
stars have $\overline{\vphi} \sim 120$ km s$^{-1}$).  Volumes 1 and
2 are shown to illustrate the difference in the distribution of stars
due to the $m=2$ deviations from axisymmetry in the spatial
distributions of disc and satellite stars. Satellite particles are
over-represented in this figure by a factor of five.}
\label{grid-toomre}
\end{center}
\end{figure}

Fig. \ref{grid-toomre} shows the Toomre diagrams for the ``$z$=1''
experiment with a prograde discy satellite with initial inclination
30$\degr$, including disc and satellite stars. Note that the satellite
particles are over-represented in this Figure by a factor five. The
three velocity components have been convolved with Gaussian errors of
10, 20 and 30 km s$^{-1}$.  The volumes contain different number of
particles due to the asymmetric spatial distributions discussed in
Section ~\ref{sec:bars}. The separation between both types of stars is
evident even if the errors in the velocity are 20 km s$^{-1}$ and
should allow the distinction between satellite stars and heated disc
stars in surveys such as RAVE \citep{steinmetz2006}.

\subsection{Statistical tests on the velocity distributions}
\label{sec:stats}

\begin{figure*}
\begin{center}
\includegraphics[width=80mm]{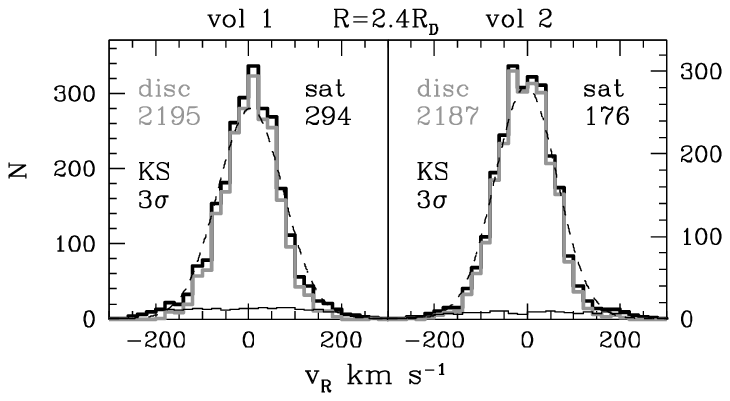}
\includegraphics[width=80mm]{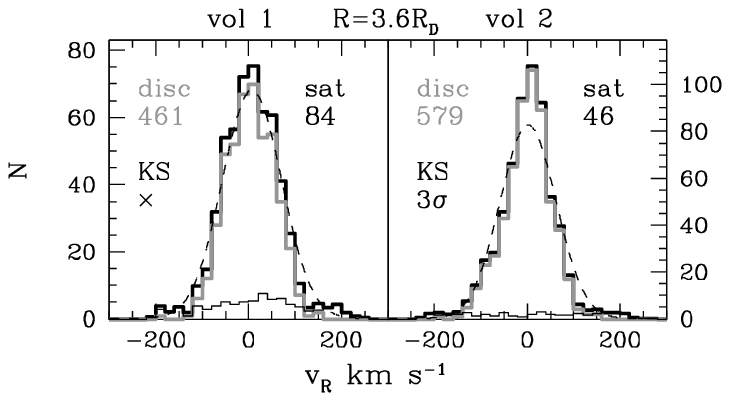}\\
\includegraphics[width=80mm]{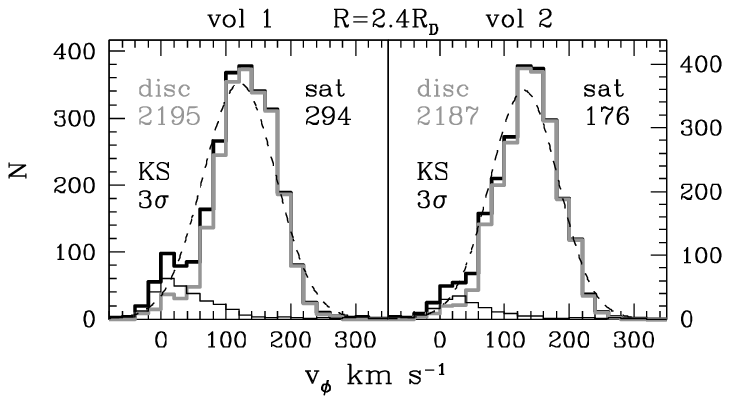}
\includegraphics[width=80mm]{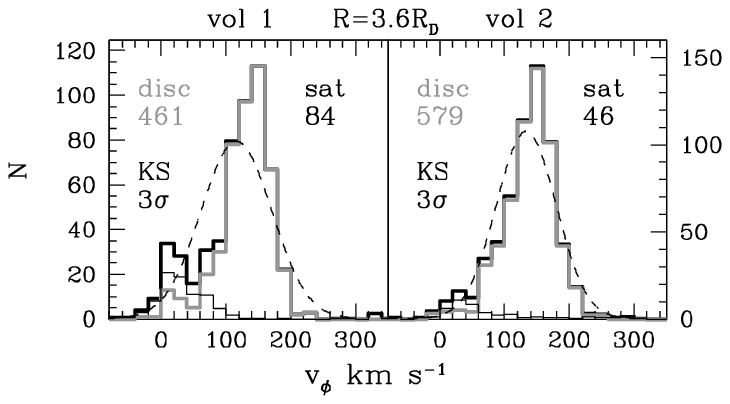}\\
\caption{ Histograms of the radial (top) and tangential (bottom)
velocities of stars within volumes centred at $R=2.4R_{\rm D}$ and
$R=3.6R_{\rm D}$, for ``$z$=1'' experiments with a prograde discy satellite
with initial orbital inclination 30$\degr$.  KS/Kuiper tests have been
performed in order to obtain a probability that the total distribution
of stars (thick black line) is drawn from a Gaussian distribution
(obtained as the best fit to the total distribution).  Distributions
of disc (grey line) and satellite (thin-black line) stars are also
shown.  The labels ``KS 3$\sigma$, 2$\sigma$ or `$\times$' (meaning
less than 1$\sigma$)'' indicate the confidence level with which one
can establish that the distribution has not been drawn from the best
fit Gaussian. Volumes 1 and 2 are shown to illustrate the difference
in the distributions due to asymmetric spatial distributions of disc
and satellite stars.  }
\label{grid-vel-histos-vrvpvz-z1}
\end{center}
\end{figure*}

\subsubsection{KS test on the $\vR$, $\vphi$ and $\vz$ distributions}
\label{vr-vp-vz-distr}

As just discussed, significant merger events like those simulated in
\citetalias{villalobos-helmi2008} leave clear and long-lasting signatures in the kinematics of
the final systems.  Fig. \ref{vrvpvz} shows that disc and satellite
stars clearly dominate different regions of velocity space, and that
the final distributions are not Gaussian.  In this section we quantify
this non-Gaussianity, by studying the 1-dimensional velocity
distributions along the $R$, $\phi$ and $z$ directions in our local
volumes. To this end we perform the
Kolmogorov-Smirnov/Kuiper\footnote{The Kuiper variant of the test is
chosen instead of the standard Kolmogorov-Smirnov since the first
guarantees equal sensitivity at all values of the distribution,
particularly in the wings.} statistical test \citep[e.g.,
see][]{press1992}.  Our null hypothesis is that the underlying
velocity distributions are Gaussian along each one of the principal
axes of the velocity ellipsoid ($\vR$, $v_\phi$, $\vz$).

Fig.~\ref{grid-vel-histos-vrvpvz-z1} shows the disc+satellite velocity
distributions of $\vR$ and $\vphi$ and the best Gaussian fit
(dashed lines) for a ``$z$=1'' experiment with a prograde discy
satellite with initial orbital inclination 30$\degr$.  The stars
considered here are located in volumes of 2 kpc radius at $R=2.4R_{\rm D}$
and $R=3.6R_{\rm D}$ (left and right panels, respectively), and the number
of satellite stars has been properly normalised (effectively, only 1
in 5 stars is considered).  This figure shows that in most cases our
null hypothesis can be rejected with better than $3\sigma$
confidence. This is mainly because the simulated velocity
distributions have rather extended wings. These wings are
preferentially dominated by satellite stars, also in the case of the
$\vz$ distributions (see Fig. \ref{vrvpvz}). However, in this latter
case the contribution of satellite stars to the wings is too low (by
number) to be detected by the test.  As expected (Section \ref{on-kine})
the results of this test are nearly independent of the azimuthal
location of the volumes. Note that even in these small samples, the
features are very significant, and should be relatively easy to detect
observationally.  Furthermore, the velocity distributions do bear a
close resemblance to those seen in the in-situ samples studied by
\citet{gil2002}.

\begin{figure}
\begin{center}
\includegraphics[width=85mm]{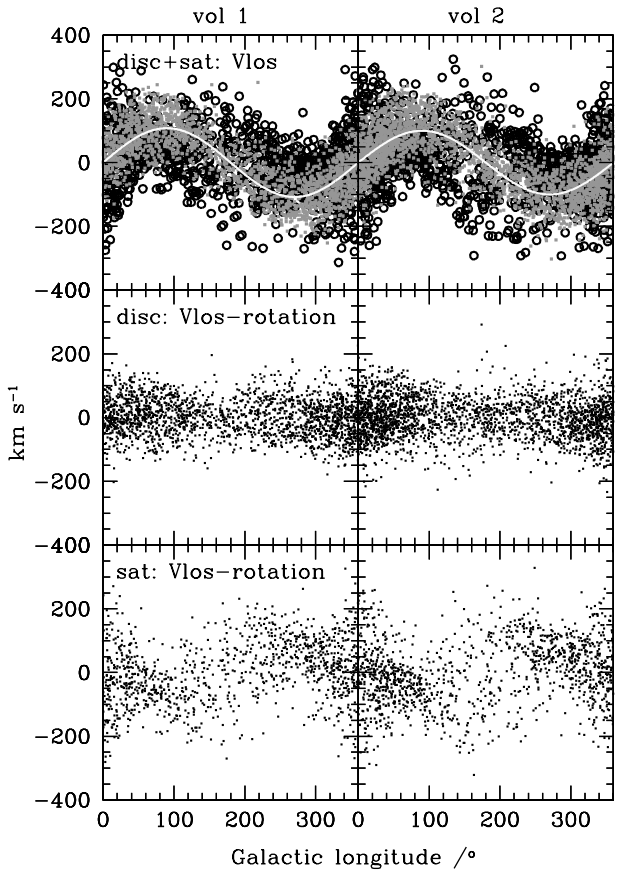}
\caption{ \emph{Top panels}: Heliocentric line-of-sight velocities of
disc (grey) and satellite (black) stars.  The white line shows a
sin($l$) fit to model the mean rotational velocity of the final thick
disc.  \emph{Middle and bottom panels}: The rotation signal has been
subtracted for both disc and satellite stars, respectively.  The
figures correspond to the ``$z$=1'' experiment using a prograde
spherical satellite with initial orbital inclination 30$\degr$.
Volumes 1 and 2 are shown to illustrate the difference in the
distributions due to asymmetric spatial distributions of disc and
satellite stars. In all the panels, the satellite stars are
over-represented by a factor five.}
\label{grid-vlos-scatter-spherical-z1}
\end{center}
\end{figure}

\subsubsection{Heliocentric line-of-sight velocities, $\vlos$}

In practice, it is relatively difficult to obtain accurate full space
velocities for large samples of stars. The main challenge being to
measure accurately their proper motions and distances for the stars.  However,
a possible workaround to this problem is to concentrate on
heliocentric line-of-sight velocities ($\vlos$) 
which are comparatively cheaper to obtain and can be measured with 
high accuracy for large samples of stars.

In our simulations we compute the $\vlos$ as follows. 
First both the galactic longitude ($l$) and latitude ($b$) 
are computed for each star with respect to the centre of the 
local volume, with the $+x$-axis pointing towards the galactic 
centre and the $+y$-axis in the sense of rotation of the system.
Then $\vlos$ is computed as:
     
\begin{equation} \label{eq-vlos}
\vlos = v_{\rm x} \cos l  \cos b  + v_{\rm y} \sin l \cos b + \vz \sin b
\end{equation}
where the three velocity components of each star are measured with
respect to the galactic centre (i.e. they include the galactic
rotation). This implies that stars on circular low latitude orbits
have $\vlos \sim v_{\rm circ} \sin l \cos b$, since $v_{\rm x}$
and $\vz$ should both be (close to) zero.

The upper panels of Fig. \ref{grid-vlos-scatter-spherical-z1} show the
$\vlos$ of disc (grey) and satellite (black) stars as a
function of their galactic longitude for the ``$z$=1'' experiment with a
prograde spherical satellite with initial orbital inclination of
30$\degr$.  The most evident feature is that stars of the heated disc
are still rotating on fairly circular orbits in spite of the
relatively massive minor merger, as demonstrated by the sinusoidal
dependence on galactic longitude of their line-of-sight velocities.

We proceed to subtract the mean rotation of the system,
$v_{\rm{rot}}$, in order to better quantify the contribution of
the satellite stars to the wings of the distributions of
$\vlos$. The mean rotation is found by fitting to the
$\vlos(l)$ distribution of both disc and satellite stars
(after the proper normalisation) the one-parameter function
$v_{\rm{rot}} \sin (l)$. The result is the white curve in the top
panel of Fig. \ref{grid-vlos-scatter-spherical-z1}.  The middle and
bottom panels of this Figure show both disc and stars radial
velocities obtained after subtracting $v_{\rm{rot}}$. The
different behaviour of disc and satellite stars is very clear now, and
this is also the case for most of our experiments. In the particular
example plotted here, the maximum separation is found around
$l\sim$100--160$\degr$ and $l\sim$220--240$\degr$ as illustrated in
the figure. Note finally that the $m=2$ deviations from axisymmetry do
not significantly impact the distributions, i.e. there are no large
volume-to-volume differences.

\begin{figure*}
\begin{center}
\includegraphics[width=59mm]{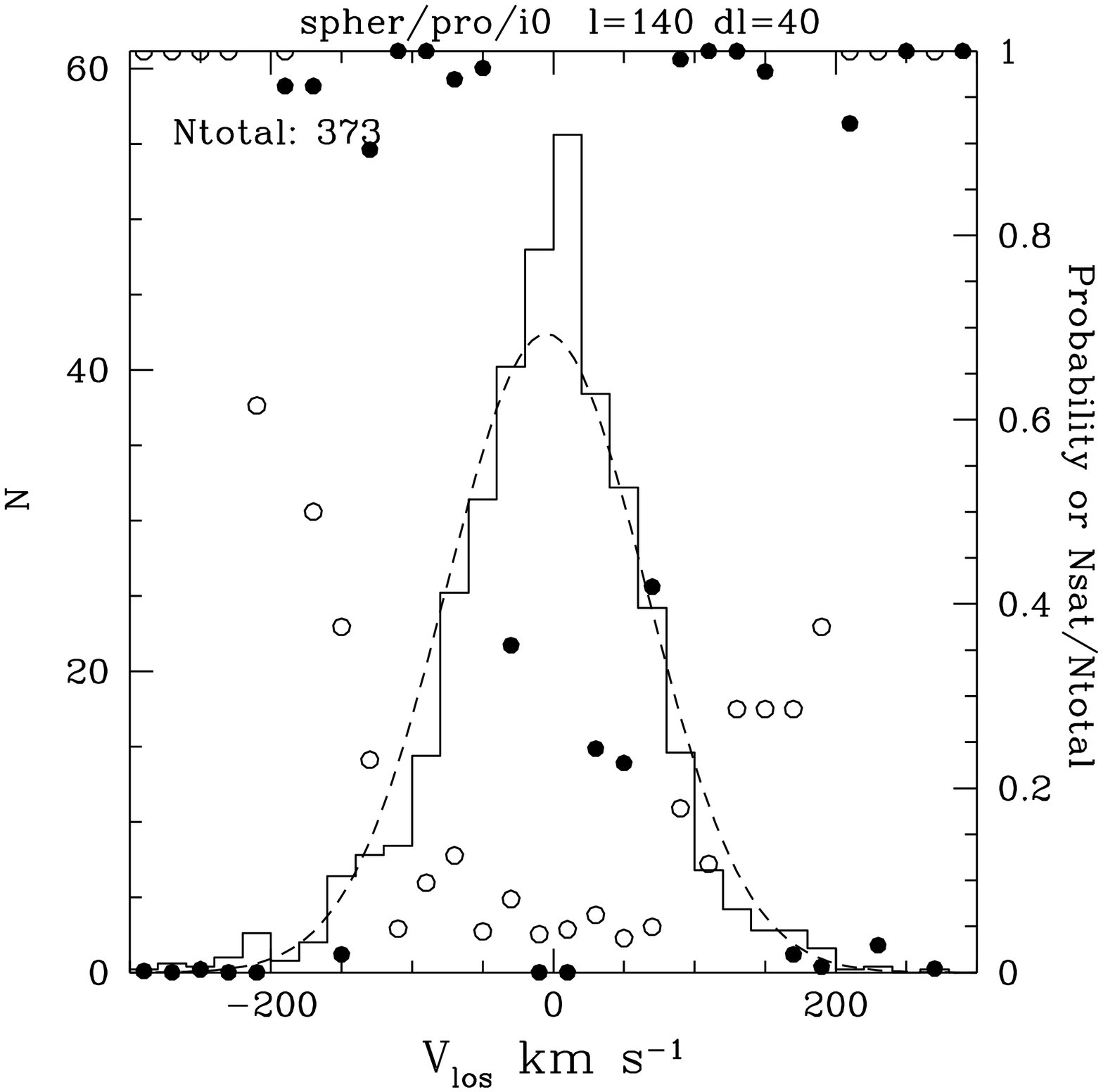}\hspace*{0.02cm}
\includegraphics[width=59mm]{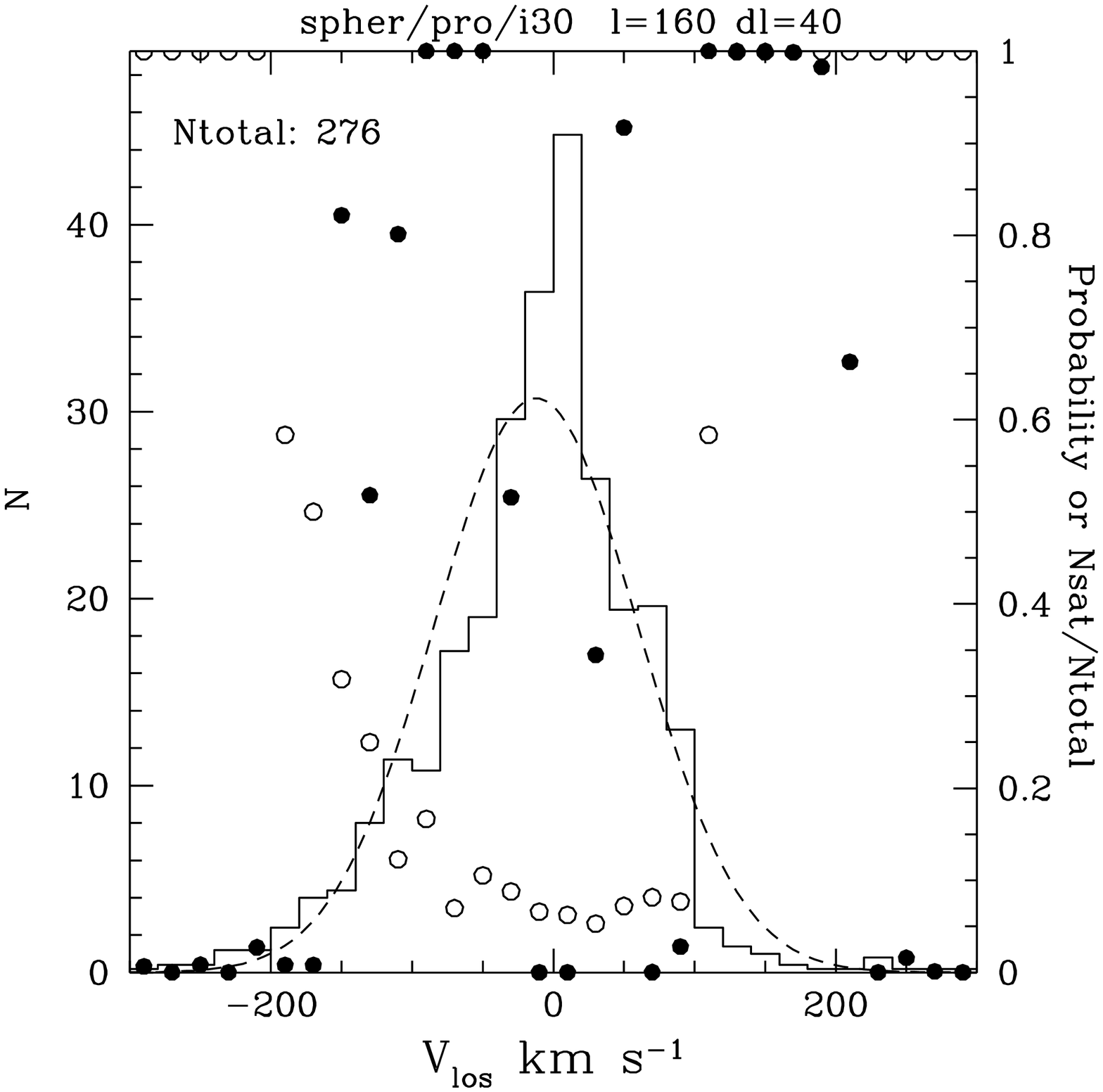}\hspace*{0.02cm}
\includegraphics[width=59mm]{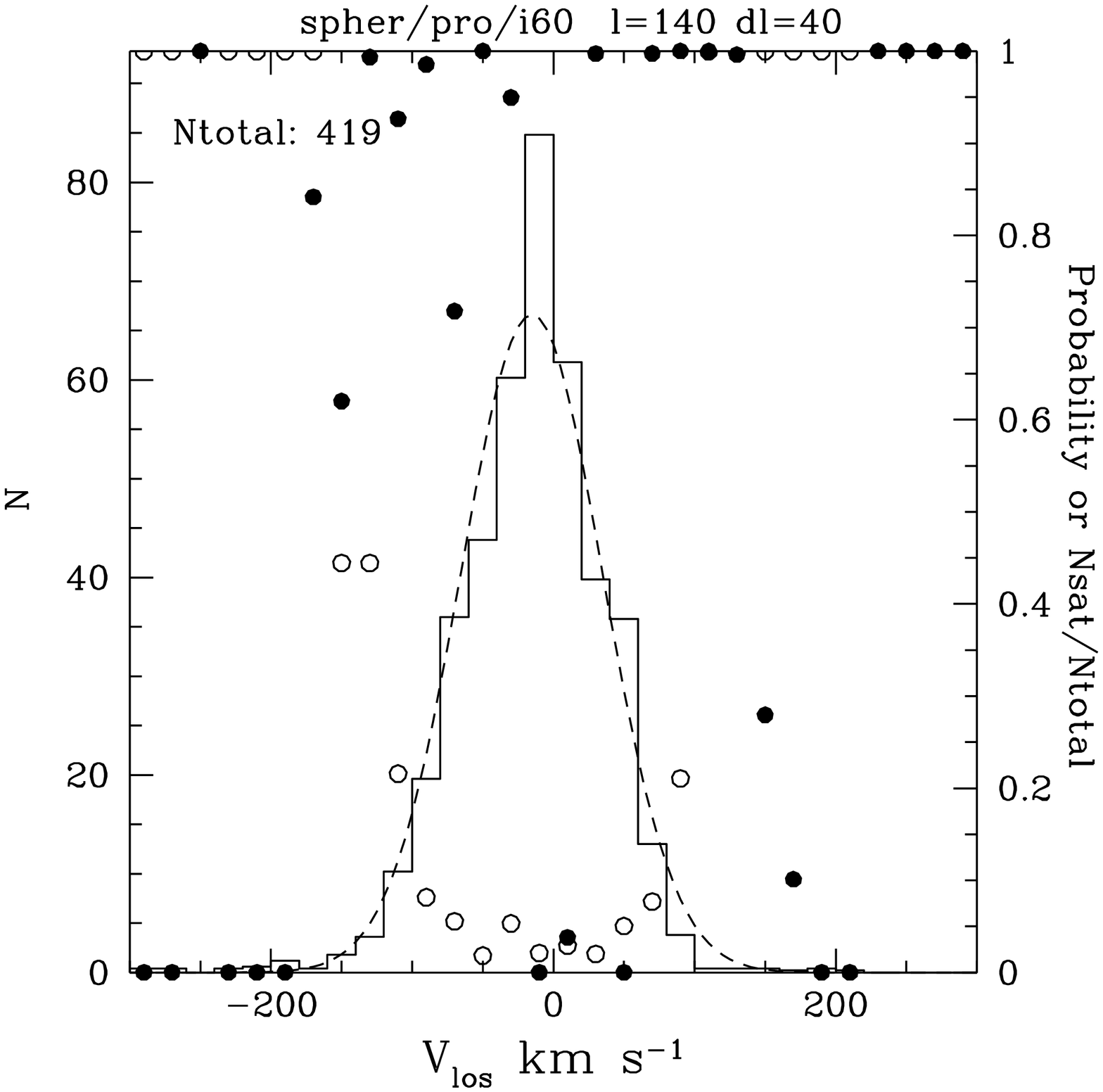}\\
\includegraphics[width=59mm]{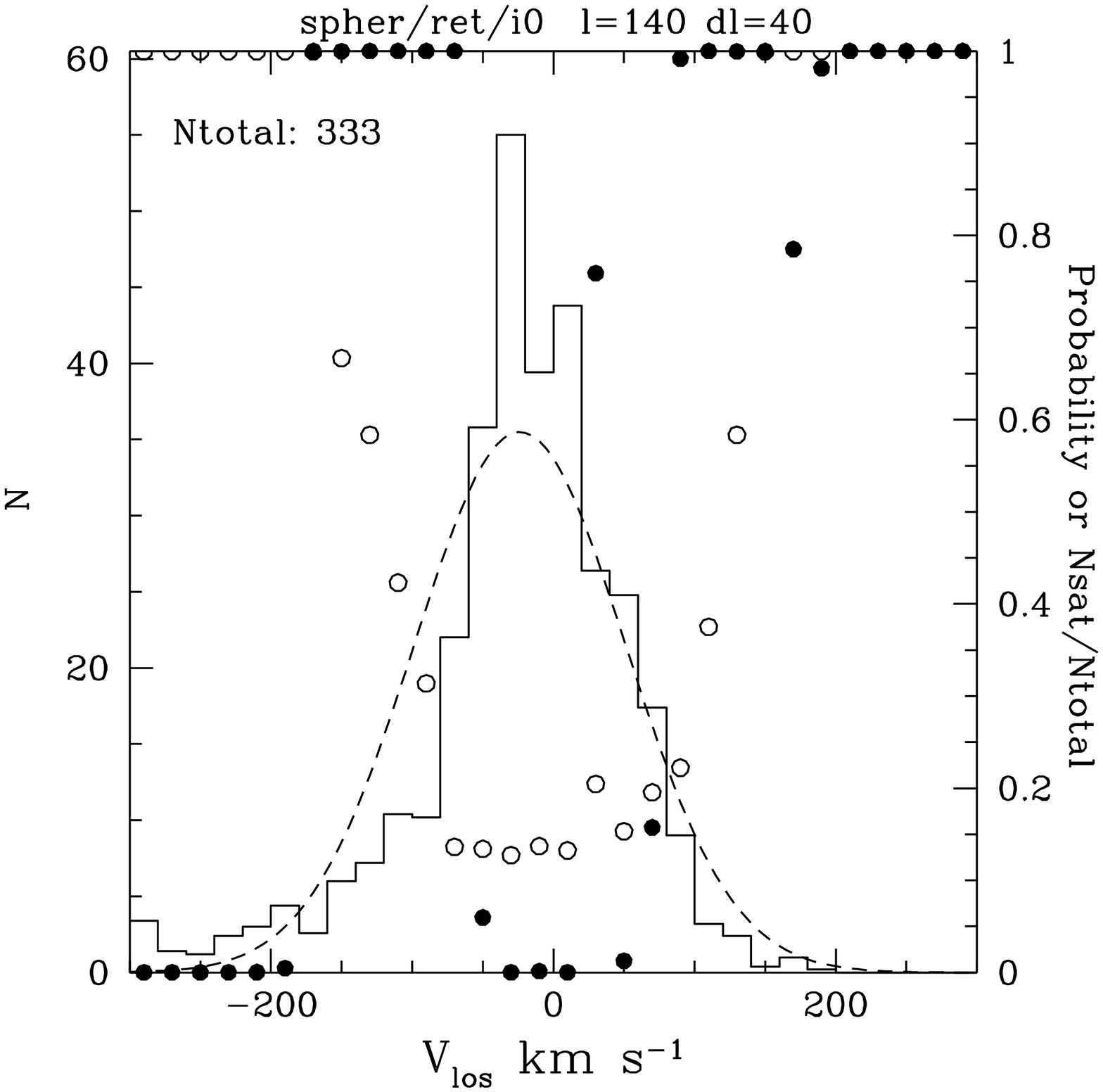}\hspace*{0.02cm}
\includegraphics[width=59mm]{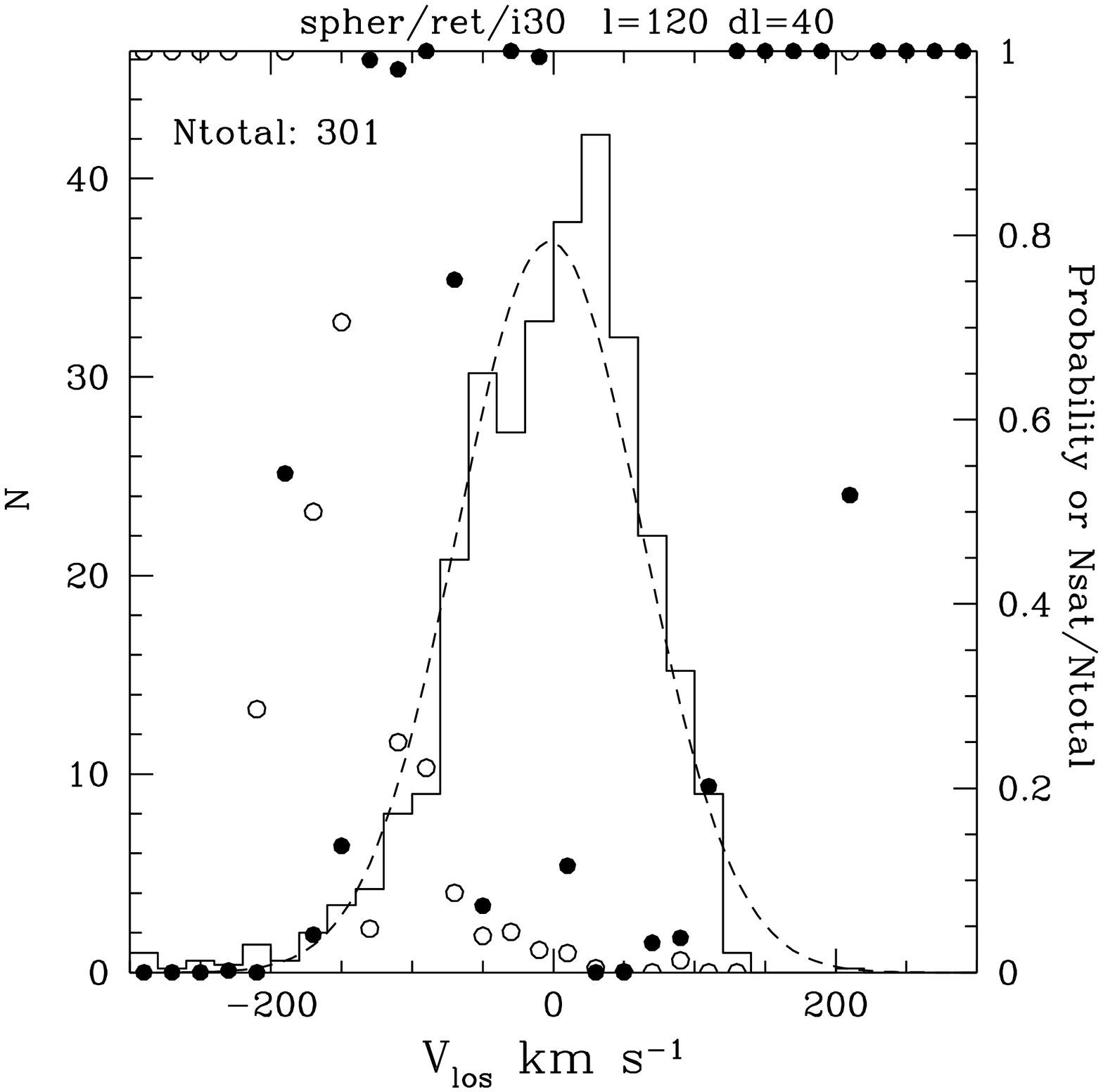}\hspace*{0.02cm}
\includegraphics[width=59mm]{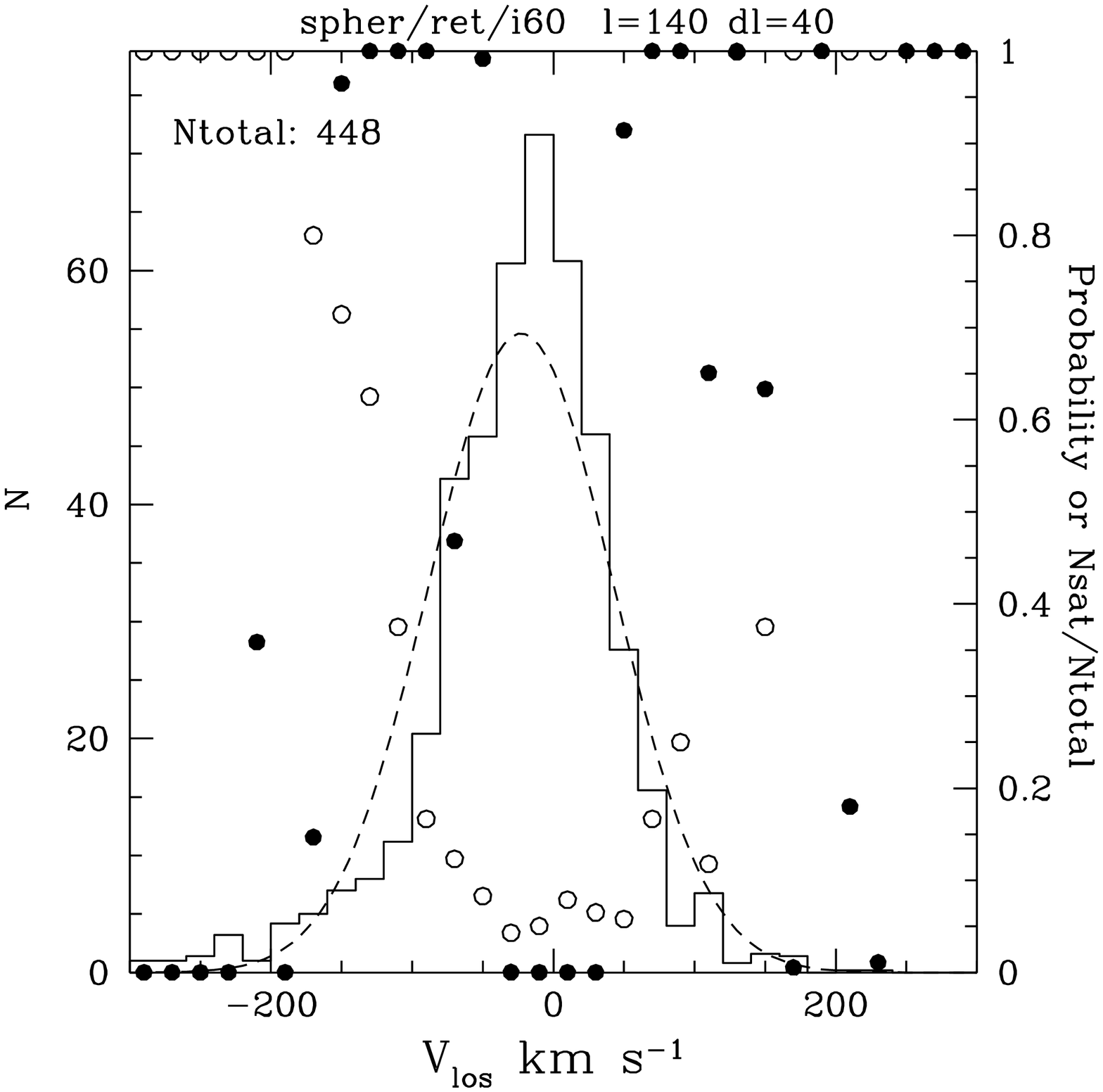}
\caption{ Histograms of the heliocentric line-of-sight velocities
after subtracting the mean rotation of the final thick disc, for
``$z$=1'' experiments using prograde and retrograde spherical satellites
with initial orbital inclinations of 0$\degr$, 30$\degr$ and
60$\degr$.  The histograms include disc and satellite stars within a
slice around $l\sim$140$\degr$, where the contribution of satellite
stars to the wings is maximal (see Fig.
\ref{grid-vlos-scatter-spherical-z1}). The width of $l$-slices is
40$\degr$.  The best Gaussian fits to the histograms are shown in
dashed.  The open circles show the fraction of satellite stars in each
velocity bin. The solid circles denote the probability of the observed
number of stars compared to what is expected from the best fit
Gaussian at each velocity bin.}
\label{grid-vlos-histos-z1}
\end{center}
\end{figure*}

In Fig. \ref{grid-vlos-histos-z1} we have plotted the
$\vlos$ distributions of disc+satellite stars, after
subtracting $v_{\rm{rot}}$, for all ``$z$=1'' experiments with
spherical satellites. The stars included in the histograms have
galactic longitudes within an interval of $\delta l=40\degr$ around
$l=140\degr$ (where the difference between the two populations of
stars was largest according to the previous Figure).  At these
longitudes, $l\sim140\degr$, most of the satellite stars have negative
$\vlos$. This is manifested in most of the distributions
shown in Fig. \ref{grid-vlos-histos-z1}, which are (slightly)
asymmetric and tend to have a prominent negative velocity tail,
particularly compared to the best fit Gaussian (denoted here by the
dashed curves). To further emphasise this point, the open circles in
Figure \ref{grid-vlos-histos-z1} denote the fraction of satellite
stars present at a given velocity (bin).

We quantify the statistical significance of the features present in
the $\vlos$ distributions by computing a probability that
measures the likelihood of the ``observed'' number of stars in a given
velocity bin. We proceed as follows. We first generate
$N_{\rm tot,real}=10^4$ random realisations based on the best Gaussian fit.
We then define the probability of observing $N_{{\rm obs,}i}$ stars or more
in the $i$-th bin as:
\begin{equation} \label{eq-prob}
P_{i}(\ge N_{{\rm obs,}i}) = \frac{N_{\rm real}(N_{{\rm gen,}i} \ge N_{{\rm obs,}i})}{N_{\rm tot,real}},
\end{equation}
where $N_{{\rm gen,}i}$ denotes the number of ``random stars'' present in the
$i$-th bin. Therefore the numerator in Eq.~(\ref{eq-prob}) denotes the
number of realisations for which the number of ``random stars'' is
greater or equal the observed one for each bin.

The resulting probabilities $P_{i}(\ge N_{{\rm obs,}i})$ are depicted as the solid
circles in Fig. \ref{grid-vlos-histos-z1}. Typically very low
probabilities ($< 1$\%) are obtained in the wings of the velocity
distribution, as well as near the central peak. Recall that the wings
are dominated by satellite stars (as indicated by the open
circles). This implies that this test is both able to identify
non-Gaussian features, as well as the presence of accreted stars. The
low probabilities found near the central peak are due to the fact that
this region is dominated by stars in a cold remnant disc, which survives the
merger event, and contains $10-20$\% of the total mass of the thick
disc (see Section 3.4 of \citetalias{villalobos-helmi2008}).

Similar results are found for all our experiments (also those not
shown in Fig. \ref{grid-vlos-histos-z1}). In general, after analysing
similar histograms for all the experiments, we find that even when the
histograms contain a relatively small number of stars, it is possible
to detect rare peaks in the wings which are composed mostly by
satellite stars. 

At this point it is important to remind the reader that the
simulations performed in \citetalias{villalobos-helmi2008} do not include the growth of a fresh
thin disc after the merger that led to the formation of the thick
disc.  This new, colder and (presumably) more massive thin disc would
\emph{mostly} dominate around $\vR \sim 0$ and $\vz \sim 0$ km
s$^{-1}$. For example, in Figure \ref{grid-vel-histos-vrvpvz-z1} it
should result in an enhancement of the central peak of the $\vR$ and
$v_{z}$ distributions.  On the other hand, in the direction of
rotation a pronounced peak should be present at
$\overline{\vphi}_{\rm ,thin}$, where $\overline{\vphi}_{\rm ,thin} >
\overline{\vphi}_{\rm ,thick}$.  This implies that the satellite stars are
expected to remain the main contributors to the wings of all the
velocity distributions.
The deposition of a significant amount of mass in the galactic plane
should lead to an increase in the rotational velocity of the thick
disc stars, and hence the line-of-sight velocities should still show a
sinusoidal dependence on longitude, but now with a larger amplitude
than that visible in Fig. \ref{grid-vlos-scatter-spherical-z1}.
Satellite stars will continue to dominate the wings of the $\vlos$
distributions and should clearly become apparent after the mean
rotation of disc has been subtracted.

\section{Summary and Conclusions}

In this follow-up study we have analysed the phase-space properties of
a sample of 24 simulated thick discs presented in \citetalias{villalobos-helmi2008}. These thick
discs have been produced via a significant merger between a
pre-existing disc galaxy and a satellite.  In that study several
combinations of the initial conditions of the progenitors were
explored, such as: two redshifts of formation (``$z$=0'' and ``$z$=1'');
two mass ratios between the main disc galaxy and the satellite (10\%
and 20\%); two different morphologies for the stellar component of the
satellite (spherical and discy); and three initial orbital inclination
of the satellites (0$\degr$, 30$\degr$ and 60$\degr$) in both prograde
and retrograde orbits.

The goal of this paper has been to find robust indicators of the
merger origin of these simulated thick discs by characterising their
phase-space structure. This involves establishing which properties
from the progenitors have been retained in the final system as well as
ways to distinguish dynamically in-situ and accreted stars. Our
ultimate goal is to shed light onto the origin of the thick disc of
the Milky Way by comparing the predictions of this model to already
available (and future) surveys of nearby stars such as RAVE, SEGUE and
{\em Gaia}.

Our simulations show that the final spatial distributions of stars
from both the heated disc and satellite are usually asymmetric with
respect to the rotation axis of the system, but that each of these
``$m=2$'' deviations are generally out of phase. The lack of apparent
correlation between these asymmetries suggests that they could
originate in different dynamical processes. Indeed, the ``bar-like''
distribution of heated disc stars seems to be induced by the
asymmetric perturbation of the decaying intruder, while the asymmetry
in the distribution of satellite stars bears some analogy to the
radial orbit instability. It is interesting to notice that similar
spatially asymmetric distributions have been observed in the thick
disc of the Milky Way \citep[e.g.][]{larsen08}. In terms of
kinematics, these asymmetries are found to have a negligible effect on
the velocity distributions of both heated disc and satellite stars.

When samples of thick disc stars are selected in small volumes that
resemble ``solar neighbourhoods'', we find clear differences in the
velocity distributions of in-situ and accreted stars. The stars from
the heated disc are more centrally concentrated in $\vR$ and $\vz$
while the accreted stars show broader velocity distributions in these
directions. Additionally, the accreted stars rotate more slowly and
they show a characteristic ``banana-shaped'' distribution on the
$\vR-\vphi$ plane. It is important to note that similar features
are observed independently of the size and location of the volumes on
the midplane.

The vertical component of the angular momentum, $\Lz$, as function of
distance from the galaxy's centre is found to be a clear discriminator
to separate heated disc stars from those accreted. In all our
experiments in-situ stars have $\Lz \propto R$ whereas for accreted
stars $\Lz$ is approximately constant (i.e. independent of $R$).  This
implies that the distribution of $\Lz$ is predicted to be bimodal, and
that the bimodal nature should become more apparent with increasing
distance from the galactic centre. The $\Lz \propto R$ behaviour thus
provides a clean test of the presence of a pre-existing disc from
which the Galactic thick disc stems.  Note that such a behaviour would
also be expected if the thick disc would have resulted from resonant
interactions with transient spiral arms as proposed by
\citet{roskar2008}. In this case, however, one would not expect to
find a second component associated to the accreted satellite.

We also find that heliocentric line-of-sight velocities ($\vlos$) as
a function of galactic longitude show that most heated disc stars
remain on nearly circular orbits. This implies that even after a
significant merger, the heated disc is able to retain part of the
dynamical characteristics of the pre-existing disc (namely the
relatively low orbital eccentricities). After subtracting the mean
rotation, the wings of the $\vlos$ distributions are found to 
contain mostly accreted stars.  Our analysis shows that the contribution of
accreted stars to the wings is statistically measurable and, in
principle, it could be easily detected by surveys in the solar
neighbourhood.

Finally, it is important to highlight the robustness of the results
presented in this paper regarding the different initial conditions
explored. For instance, the significant separation between heated disc
stars and satellite stars in terms of $\Lz$ as a function of galactic
radius.  Even though the separation shows a mild trend with the
satellite's initial orbital inclination, in all cases the separation
is clear, especially when stars are located at large galactocentric
radii.  Similarly, the sinusoidal behaviour of $\vlos(l)$ is
observed in all experiments independently of the size or location of
the volumes.  The robustness of these results allow us to consider
them as direct probes of the disc heating scenario for the formation
of thick discs.  We aim to soon test these predictions on surveys of
nearby stars such as RAVE to shed light on the formation of the
Galactic thick disc.

\section*{Acknowledgements}
We acknowledge financial support from from the Netherlands
Organisation for Scientific Research (NWO). The simulations were run
in the Linux cluster at the Centre for High Performance Computing and
Visualisation (HPC/V) of the University of Groningen in The
Netherlands.

\bibliographystyle{mn}
\bibliography{bibl_thickdisc2} 

%\label{lastpage}
\end{document}